\documentclass[letter,longauth]{aa}  

\usepackage{graphicx}
\usepackage{txfonts}
\usepackage{subfigure}
\usepackage[colorinlistoftodos]{todonotes}
\usepackage{color, colortbl}
\usepackage[colorlinks=true, allcolors=blue]{hyperref}
\definecolor{Red}{rgb}{0.9,0,0}
\definecolor{Blue}{rgb}{0,0,0.9}
\usepackage{natbib}
\bibpunct{(}{)}{;}{a}{}{,} 
\usepackage{ulem}
\usepackage{amstext}

\begin{document}

   \title{First stellar occultation by the Galilean moon Europa and upcoming events between 2019 and 2021}
   \titlerunning{Stellar occultation by Europa}
   \authorrunning{B. Morgado, G. Benedetti-Rossi, A. R. Gomes-J\'unior et al.}
   \author{B. Morgado
          \inst{1,2}\fnmsep\thanks{\email{morgado.fis@gmail.com}},
          G. Benedetti-Rossi\inst{1,2}, 
          A. R. Gomes-J\'unior\inst{1,2,3,4},
          M. Assafin\inst{2,3},
          V. Lainey\inst{5,6},
          R. Vieira-Martins\inst{1,2,3},
          J. I. B. Camargo\inst{1,2},
          F. Braga-Ribas\inst{7,2,8},
          R. C. Boufleur\inst{1,2}, 
          J. Fabrega\inst{9},
          D. I. Machado\inst{10,11},
          A. Maury\inst{9},\\
          L. L. Trabuco\inst{11},
          J. R. de Barros\inst{12},
          P. Cacella\inst{13},
          A. Crispim\inst{7},
          C. Jaques\inst{12},
          G. Y. Navas\inst{14},
          E. Pimentel\inst{12},\\
          F. L. Rommel\inst{1,2,7},
          T. de Santana\inst{4},
          W. Schoenell\inst{15},
          R. Sfair\inst{4}
          \and
          O. C. Winter\inst{4}
          }

   \institute{
   Observat\'orio Nacional/MCTIC, R. General Jos\'e Cristino 77, Rio de Janeiro, RJ 20.921-400, Brazil
   \and
   Laborat\'orio Interinstitucional de e-Astronomia - LIneA \& INCT do e-Universo, Rua Gal. Jos\'e Cristino 77, Rio de Janeiro, RJ 20921-400, Brazil
   \and
   Observat\'orio do Valongo/UFRJ, Ladeira Pedro Antonio 43, Rio de Janeiro, RJ 20080-090, Brazil
   \and
   UNESP - S\~ao Paulo State University, SP 12516-410, Guaratinguet\'a, S\~ao Paulo, Brazil
   \and
   Jet Propulsion Laboratory, California Institute of Technology, 4800 Oak Grove Drive, Pasadena, CA 91109-8099, United States
   \and
   IMCCE, Observatoire de Paris, PSL Research University, CNRS-UMR 8028, Sorbonne Universit\'es, UPMC, Univ. Lille 1, 77 \\Av. Denfert-Rochereau, 75014 Paris, France
   \and
   Federal University of Technology - Paran\'a (UTFPR/DAFIS), Av. Sete de Setembro, 3165, CEP 80230-901, Paran\'a, Brazil
   \and
   LESIA, Observatoire de Paris, Universit\'e PSL, CNRS, Sorbonne Universit\'e, Univ. Paris Diderot, Sorbonne Paris Cit\'e, 5 place Jules Janssen, 92195 Meudon, France  
   \and
   San Pedro de Atacama Celestial Explorations, Casilla 21, San Pedro de Atacama, Chile
   \and
   Unioeste, Avenida Tarqu\'inio Joslin dos Santos 1300, Foz do Igua\c{c}u, PR 85870-650, Paran\'a, Brazil
   \and
   Polo Astron\^omico Casimiro Montenegro Filho/FPTI-BR, Avenida Tancredo Neves 6731, Foz do Igua\c{c}u, PR 85867-900, Paran\'a, Brazil
   \and
   Observat\'orio SONEAR, Brazil
   \and
   Dogsheaven Observatory, SMPW Q25 CJ1 LT10B, Brasilia, Brazil
   \and
   Research Center of Astronomy, Francisco J. Duarte (CIDA). 264, 5101-A Mérida, Venezuela.
   \and
   Instituto de F\'isica/UFRGS, Porto Alegre, RS 91501-970, Brazil, 
       }
   \date{Received XX; accepted XX}

 
  \abstract
   {Bright stellar positions are now known with an uncertainty below 1 mas thanks to Gaia DR2. Between 2019-2020, the Galactic plane will be the background of Jupiter. The dense stellar background will lead to an increase in the number of occultations, while the Gaia DR2 catalogue will reduce the prediction uncertainties for the shadow path.}
   {We observed a stellar occultation by the Galilean moon Europa (J2) and propose a campaign for observing stellar occultations for all Galilean moons.}
   {During a predicted period of time, we measured the light flux of the occulted star and the object to determine the time when the flux dropped with respect to one or more reference stars, and the time that it rose again for each observational station. The chords obtained from these observations allowed us to determine apparent sizes, oblatness, and positions with kilometre accuracy.}
   {We present results obtained from the first stellar occultation by the Galilean moon Europa observed on 2017 March 31. The apparent fitted ellipse presents an equivalent radius of 1561.2 $\pm$ 3.6 km and oblatenesses 0.0010 $\pm$ 0.0028. A very precise Europa position was determined with an uncertainty of 0.8 mas. We also present prospects for a campaign to observe the future events that will occur between 2019 and 2021 for all Galilean moons.}
  {Stellar occultation is a suitable technique for obtaining physical parameters and highly accurate positions of bright satellites close to their primary. A number of successful events can render the 3D shapes of the Galilean moons with high accuracy. We encourage the observational community (amateurs included) to observe the future predicted events.}

   \maketitle 
%

\section{Introduction}

A stellar occultation occurs when a solar system object passes in front of a star from the point of view of an observer on Earth, causing a temporary drop in the observed flux of the star. This technique allows the determination of sizes and shapes with kilometre precision and to obtain characteristics of the object, such as its albedo, the presence of an atmosphere, rings, jets, or other structures around the body \citep{Sicardy2011,Sicardy2016,Braga-Ribas2013,Braga-Ribas2014,Dias-Oliveira2015,Benedetti-Rossi2016, Ortiz2015,Ortiz2017,Leiva2017,Berard2017}, or even the detection of topographic features \citep{Dias-Oliveira2017}.

Stellar occultations can also provide very accurate astrometric measurements of the occulted body, with uncertainties that can be as low as 5 to 10 km or even smaller for some objects \citep{Leiva2017,Desmars2019}. Compared with other methods in the context of the Galilean moons, classical CCD astrometry enables us to obtain positions with uncertainties in the 300-450 km level \citep{Kiseleva2008}, and relative positions between two close satellites achieve uncertainties of 100 km \citep{Peng2012}. Positions obtained using mutual phenomena have uncertainties at a level of 15-60 km \citep{Saquet2018,Arlot2014,Dias-Oliveira2013,Emelyanov2009}, but they only occur at every equinox of the host planet (every six years in the case of Jupiter). The technique of mutual approximations also provides positions with uncertainties between 15-60 km, and this does not depend on the equinox of Jupiter (see \cite{Morgado2016,Morgado2019} for details). Stellar occultation is then the only ground-based technique that can furnish astrometric measurements that are comparable with space probes, which usually have uncertainties smaller than 5 km \citep{Tajeddine2015}. In addition, the positions and sizes that can be obtained with stellar occultations are independent of reflectance models, which may have systematic errors due to variations on the satellite surface \citep{Lindegren1977}.

In the context of the Galilean satellites, stellar occultations can provide shapes and sizes with uncertainties that are comparable with those of space probe images, but that are not affected by albedo variations on the satellite surface or by limb fitting, which is highly affected by the solar phase angle. From an astrometric point of view, these events can provide the best ground-based astrometry of these moons, with uncertainties smaller than 1 mas. This is at least one order of magnitude better than other methods.

Between 2019 and 2020, Jupiter will be in a very dense star region, the Galactic centre will be its background. This will only occur again in 2031. The probability of a stellar occultation by the Jovian moons increases dramatically \citep{Gomes-Junior2016}. This is a great opportunity to observe stellar occultations by the Galilean moons, determine their positions, improve their ephemerides, and measure their shapes independently of probes. Accurate orbits help to prepare space missions targeting the Jovian system \citep{Dirkx2016,Dirkx2017}, and can help in the study of tides \citep{Lainey2009,Lainey2016}. In the near future, we will have the ESA JUICE\footnote{Website: \url{http://sci.esa.int/juice/}} and the NASA \textit{Europa Clipper}\footnote{Website: \url{https://www.nasa.gov/europa}} missions, which are scheduled for launch in the next decade (2020s). Moreover, only two stellar occultations by Ganymede have been observed in the past: one in 1972 \citep{Carlson1973} and another in 2016 \citep{Daversa2017}. No occultations by Io, Europa, or Callisto are reported. Here we present the results obtained by the first observation of a stellar occultation by Europa, which occurred on 2017 March 31. These results serve as a proof of concept for the 2019-2021 campaign that is being organised.

The prediction method and the observational campaign are outlined in Section \ref{Sec:obs}, while the data analysis and the occultation results are given in Section \ref{Sec:result}. The upcoming events for the period 2019 to 2021 can be found in Section \ref{Sec:future}. Our final remarks are contained in Section \ref{Sec:conc}.

\section{Prediction and observational campaign}\label{Sec:obs}

The passage of Jupiter in a region that has the Galactic plane as background, as seen from Earth in the period between 2019 -- 2020, creates a number of opportunities for observing stellar occultations by its satellites because this region has a high density of stars \citep{Gomes-Junior2016}. This passage motivates the search for events involving the Galilean satellites because the satellites need stars with a magnitude of G = 11.5 or lower, so that the magnitude drop ($\Delta mag$) can be higher than 0.005 mag. This drop is otherwise very hard to observe with current equipment and techniques. 

Following similar procedures as described in \cite{Assafin2010,Assafin2012}, we predicted that Europa would occult a star (mag. G = 9.5) on 2017 March 31 at 06:44 UTC. The shadow path would be crossing South America with a velocity of 17.78 km/s. The prediction was made using the Europa ephemerides \textit{jup310}\footnote{Website: \url{https://naif.jpl.nasa.gov/naif/toolkit.html}} with DE435 furnished by the Jet Propulsion Laboratory (JPL), and the stellar position was obtained from the Gaia DR1 catalogue \citep{Gaia2016} because the occultation occurred before the launch of Gaia DR2 in April 2018 \citep{Gaia2018}. The star was then identified in Gaia DR2. Its position (right ascension $\alpha$, \text{declination }  - $\delta,$  and errors) was updated using its proper motion and parallax for the occultation time (2017 March 17 06:44:00 UTC), and its G mag,

\begin{eqnarray}
\alpha_{star} &=& \phantom{-}13^h~12'~15''.5430~\pm~0.16~mas, \nonumber \\
\delta_{star} &=& -05^{\circ}~56'~48''.7526~\pm~0.12~mas, \nonumber\\
Gmag &=& 9.5065. \label{Eq:prev}
\end{eqnarray}

\begin{figure}
\includegraphics[width=0.5\textwidth]{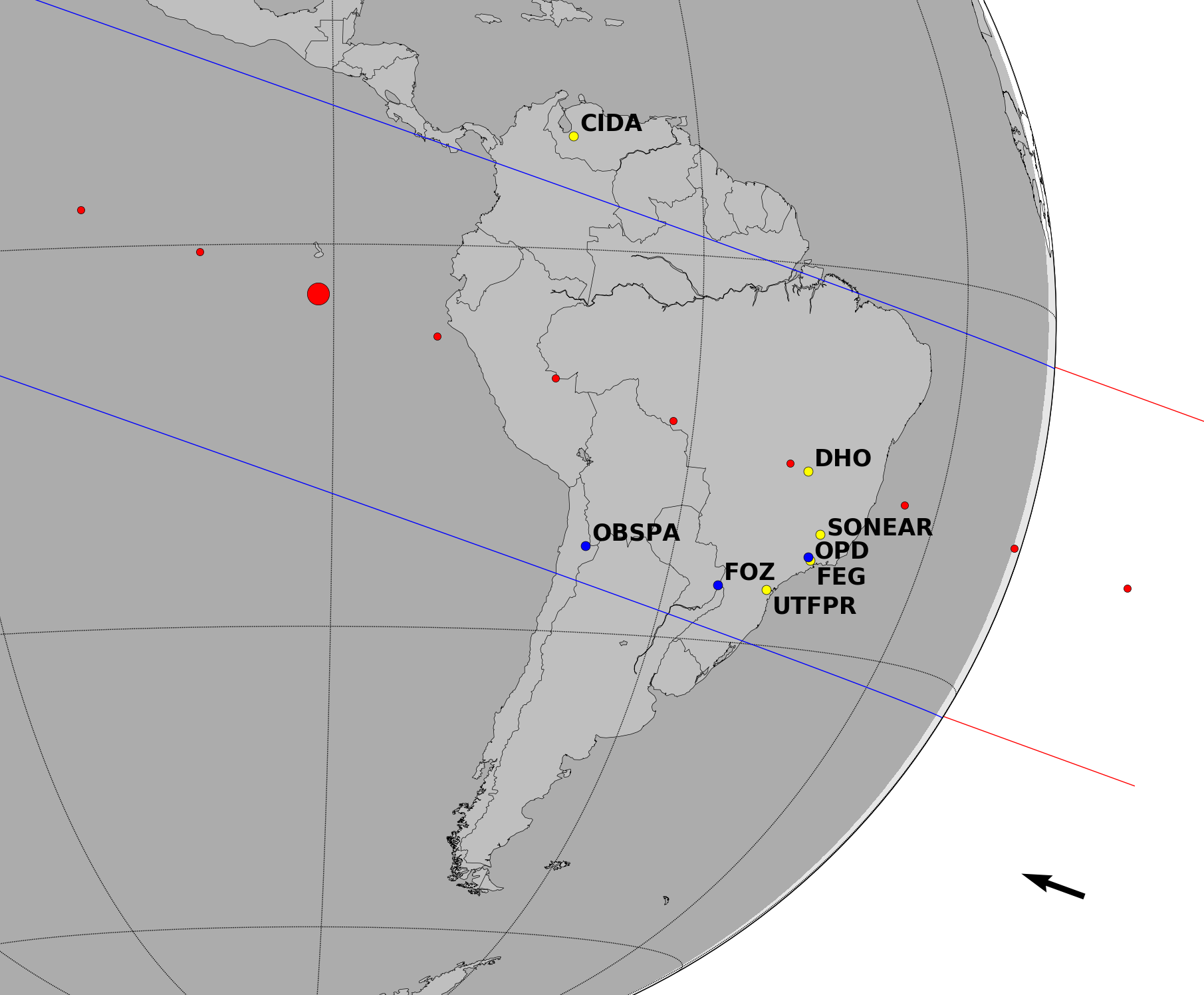}
\caption{Occultation by Europa on 2017 March 31. The centre of the body at the time of closest approach between the shadow and the geocentre (CA), at 06:44:36  UTC,  is represented by the large red dot. Blue lines represent the size limit of Europa (1561.2 $\pm$ 3.6 km). The small red dots represent the centre of the body, each separated by 1 minute from the reference instant CA. The blue dots are the observers with positive detection, and the yellow dots are observers that were unable to obtain the light curve due to bad weather conditions. The offset between our result and the prediction (using the JPL \textit{jup310} ephemeris) was -0.90 $\pm$ 3.1 km for $f_c$ and -12.3 $\pm$ 2.0 km for $g_c$, see Sect. \ref{Sec:result}. The black arrow at the bottom represents the direction of motion of the shadow.}
\label{Fig:map}
\end{figure}

This occultation was favourable for several potential observers in several countries in South America (Fig. \ref{Fig:map}). We then organised a campaign to observe this event, and a total of nine stations tried to observe the occultation. They are listed in Table \ref{tb:obs_sites}. However, due to bad weather conditions, only four were able to obtain data (OPD/B\&C, OPD/ROB, FOZ, and OBSPA)\footnote{Alias, defined in Table \ref{tb:obs_sites}.}. The observation made at Itajub\'a with the 40 cm telescope was discarded because the signal-to-noise ratio (S/N) was very low, while the other three (OPD/B\&C, FOZ, and OBSPA) obtained positive detection.     


\begin{table*}
\begin{center}
\caption{Observational stations, technical details, and circumstance.}
\begin{tabular}{lccccc}
\hline
\hline
Site   & Longitude (E) & Observers & Telescope aperture & Exposure Time & Light-curve \\
Alias  & Latitude (N)  &           & CCD                & Cycle         & rms \\
Status & Altitude (m)  &           & Filter             & (s)           & \\
\hline
\hline

Itajub\'a/MG-Brazil  & -45$^o$ 34' 57.5"& G. Benedetti-Rossi & 60 cm         & 0.60 & 0.013 \\
OPD/B\&C             & -22$^o$ 32' 07.8"& M. Assafin         & Andor/IXon-EM & 0.63 & \\
\textbf{Positive Detection}                 & 1864 &       & Methane$^1$ &  \\ 
\hline

Foz do Igua\c{c}u/PR-Brazil& -54$^o$ 35' 37.0"& D. I. Machado      & 28 cm   & 1.00 & 0.016 \\
FOZ             & -25$^o$ 26' 05.0"& L. L. Trabuco      & Raptor/Merlin  &  1.00  \\
\textbf{Positive Detection}       &184          &           & Clear &  \\
\hline

San Pedro de Atacama/Chile& -68$^o$ 10' 48.0"& A. Maury      & 40 cm   & 0.40 & 0.017 \\
OBSPA             & -22$^o$ 57' 08.0"&   J. Fabrega        & ProLine/PL16803  & 1.41   \\
\textbf{Positive Detection}          &2397          &                 & Red$^2$ &  \\
\hline

M\'erida/Venezuela  & -70$^o$ 52' 21.6"& G. R. Navas       & 100 cm   & -- & -- \\
CIDA             & +08$^o$ 47' 25.8"&    & ProLine/PL4240    &   \\
Weather Overcast      &26          &            & Red$^2$ &  \\
\hline

Bras\'ilia/DF-Brazil     & -47$^o$ 54' 39.9"& P. Cacella     & 50 cm   & -- & -- \\
DHO             & -15$^o$ 53' 29.9"&             & ASI174MM  &    \\
Weather Overcast       &1064           &                    &  --  &  \\
\hline

Itajub\'a/MG-Brazil      & -45$^o$ 34' 57.5"& W. Schoenell   & 40 cm   & -- & -- \\
OPD/ROB            & -22$^o$ 32' 07.8"&    & ASCOM/KAF16803    &    \\
Low S/N data            &1864  &        &  Red$^2$   &  \\ 
\hline

Guaratinguet\'a/SP-Brazil& -45$^o$ 11' 25.5"& R. Sfair           & 40 cm   & -- & --\\
FEG             & -22$^o$ 48' 05.5"&    T. de Santana    & Raptor/Merlin  &    \\
Weather Overcast     &543          &    O. C. Winter & -- &  \\
\hline

Curitiba/PR-Brazil     & -49$^o$ 11' 45.8"& F. Braga-Ribas     & 25 cm   & -- & --\\
OACEP             & -25$^o$ 28' 24.6"& A. Crispim         & Watec/910HX  &    \\
Weather Overcast       &861           &   F. Rommel     &  Clear  &  \\
\hline

Oliveira/MG-Brasil & -43$^o$ 59' 03.1"& C. Jaques       & 45 cm   & -- & -- \\
SONEAR             & -19$^o$ 52' 55.0"&     E. Pimentel    & ML FLI16803    &    \\
Weather Overcast        &982          &    J. R. de Barros    &  --  &  \\

\hline
\hline
\multicolumn{6}{l}{$^1$ Methane filter is a narrow-band filter centred on 889 nm and $\Delta \lambda$ = 15 nm}\\
\multicolumn{6}{l}{$^2$ Red filter from the Johnson system.}
\label{tb:obs_sites}
\end{tabular}
\end{center}

\end{table*}

Fig. \ref{Fig:map} presents the post-occultation map. It shows the shadow path of Europa through South America, and the positions of the observation stations. The blue lines represent the size of Europa obtained after the fit (See Section \ref{Sec:result}). 

\section{Data analysis and results}\label{Sec:result}

Images from all data sets were corrected for bias and flat-field using standard \textsc{iraf}\footnote{Website: \url{http://iraf.noao.edu/}} procedures. A stacking image procedure\footnote{More details in Appendix \ref{App:stack}} developed by us in \textsc{python} \citep{astropy2013}, following similar procedures as \cite{Zhu2018}, was used in FOZ images in order to obtain an S/N that was high enough to emphasise the magnitude drop. The low time-resolution at OPD/ROB (cycle $\approx$ 7 s) did not allow the use of this technique, and it was not possible to determine the magnitude drop.

\begin{figure}
\centering
\includegraphics[width=0.50\textwidth]{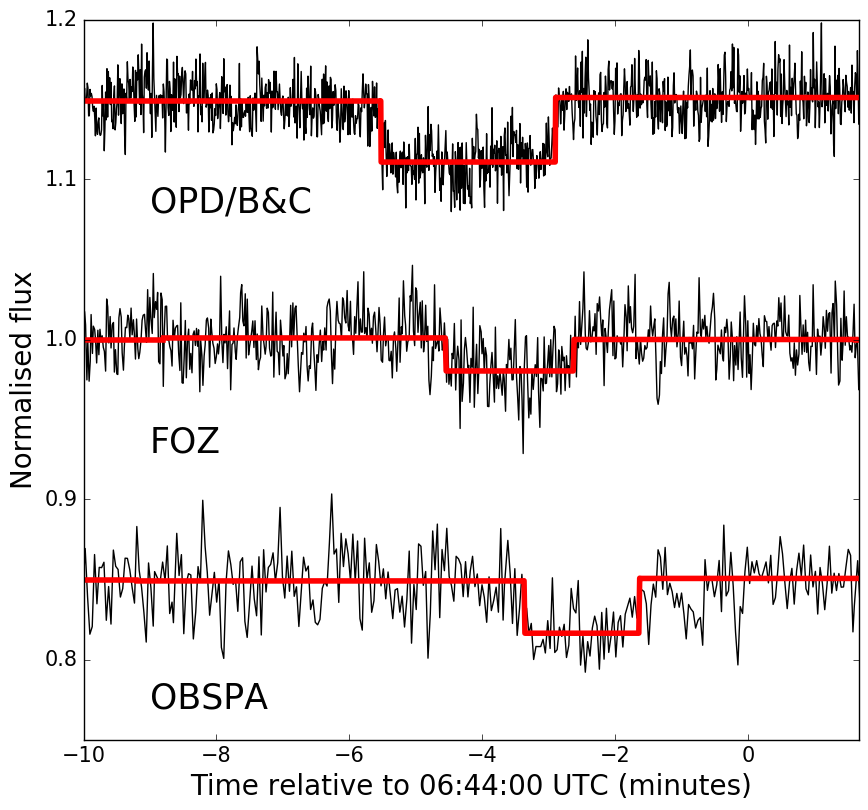}
\caption{Three normalised light curves obtained with positive detection. They are shifted vertically for better viewing. Observational circumstances are given in Table \ref{tb:obs_sites}. The difference in the depth of the curves is due to the use of different filters.}
\label{Fig:lc}
\end{figure}

Differential aperture photometry was applied with the \textsc{praia} package \citep{Assafin2011}. The calibrator and target apertures were automatically chosen to maximise the S/N for each frame. Io (J1) was set as calibrator in all sets of images because there were no other star in the field of view with a good S/N that could be used for callibration. The light fluxes of Europa and of the occulted star were measured together in the same aperture. The normalised light curves are displayed in Fig. \ref{Fig:lc}, where the black lines are the observations and the red lines are the modelled fitted light curves. 

The ingress ($t_i$) and egress ($t_e$) times were determined with a standard $\chi^2$ procedure between the observational light curve and the model. The model considers a sharp-edge occultation model convolved with Fresnel diffraction, stellar diameter (projected at the body distance, 0.633 km in our case \citep{Bourges2017}), CCD bandwidth, and finite integration time (more details in \cite{Braga-Ribas2013}). Table \ref{tb:times} contains the ingress and egress UTC times and the errors for each station with positive detection. Their respective uncertainties are given in seconds and kilometres (calculated using the event velocity of 17.78 km/s). More information about the light-curve fit procedure is presented in Appendix \ref{App:fitting}.

\begin{table}
\begin{center}
\caption{Occultation ingress (t$_i$) and egress (t$_e$) times.}
\begin{tabular}{ccc}
\hline
\hline
Station & $t_i$ & $t_e$  \\
    & (hh mm ss.ss $\pm$ s) & (hh mm ss.ss $\pm$ s) \\
    & (km) & (km) \\
\hline
OPD/B\&C & 06:38:28.55 $\pm$ 0.75 & 06:41:06.15 $\pm$ 0.24 \\
         & (13.3) & (04.3) \\
FOZ      & 06:39:26.07 $\pm$ 0.77 & 06:41:21.59 $\pm$ 1.07 \\
         & (13.7) & (19.0) \\
OBSPA    & 06:40:38.19 $\pm$ 1.20 & 06:42:22.85 $\pm$ 0.34 \\
         & (21.3) & (06.0) \\
\hline
\hline
\label{tb:times}
\end{tabular}
\end{center}
\end{table}

\begin{figure}
\includegraphics[width=0.5\textwidth]{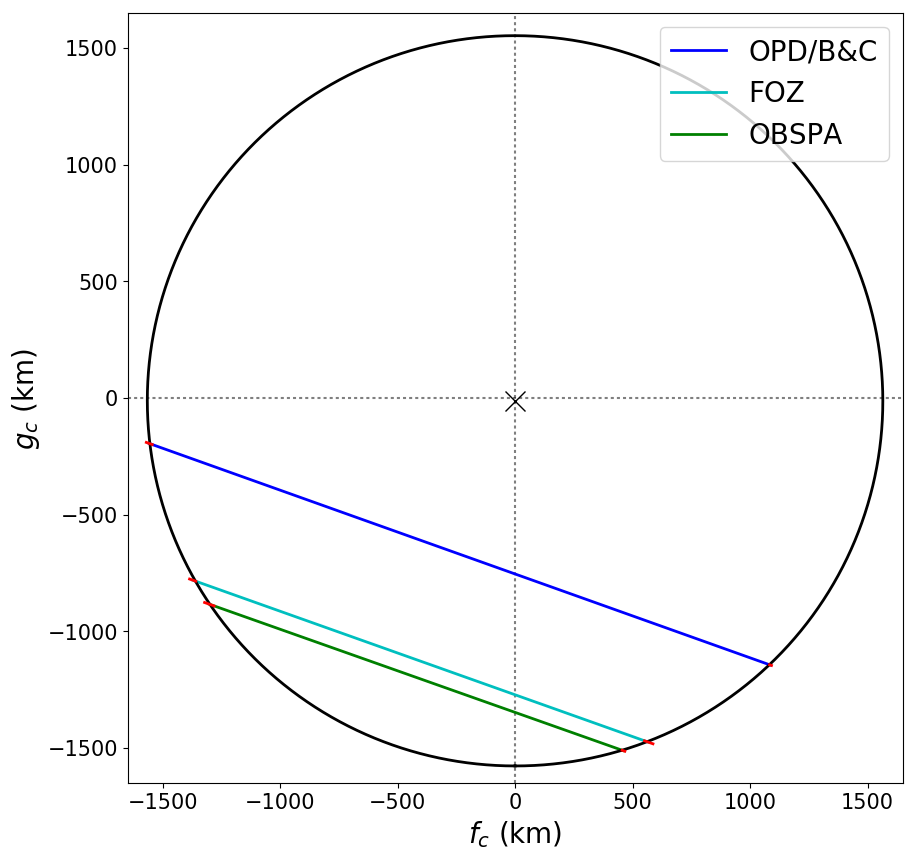}
\caption{Best elliptical limb fit (in black) using the three occultation chords. Blue, cyan, and green lines are the chords from the positive detection obtained at OPD/B\&C, FOZ, and OBSPA, respectively. The small red segments at the end of each segment indicate the 1$\sigma$ uncertainty on each chord extremity, derived from the time uncertainties provided in Table \ref{tb:times}. The position of the centre of the fitted ellipse is showed as an X, with values of $f_c =  -0.9~\pm~3.1$ km and $g_c = -12.3~\pm~2.0$ km, and represents the offset with respect to the JPL geocentric ephemeris. The fitted parameters of the ellipse are listed in Table 2.}
\label{Fig:fit}
\end{figure}

Each of the ingress and egress times were converted into a star position $(f, g)$ regarding the body centre, with $f$ and $g$ being measured positively toward local celestial east and celestial north, respectively. Each pair of positions, from the same site, is a chord, and each position is a point of which we can fit the five parameters that defines an ellipse: (i and ii) the ellipse centre ($f_c$, $g_c$), (iii) apparent semi-major axis ($a'$), (iv) the apparent oblateness ($\epsilon' = (a' - b')/a'$ , where $b'$ is the apparent semi-minor axis), and (v) the position angle of the pole $P_p$ of $b'$ . This position angle was fixed as $24.2534^{\circ}$, derived from the pole position reported by \cite{Archinal2018}. In Fig. \ref{Fig:fit} we show the three chords obtained from the occultation and the fitted ellipse. The ellipse parameters are

\begin{eqnarray}
f_c &=&  -\phantom{0}0.9~\pm~3.1~km, \nonumber\\
g_c &=&  -12.3~\pm~2.0~km, \nonumber\\
a' &=&  1562.0~\pm~3.6~km, \nonumber\\
\epsilon' &=& 0.0010~\pm~0.0028, \nonumber\\
P_p &=& 24.2534^{\circ}. \label{Eq:fit}
\end{eqnarray}

It is important to highlight that the centre position ($f_{c},g_{c}$) and the apparent size and shape of the ellipse ($a', \epsilon'$) are correlated. This correlation decreases with the number of obtained chords, and moreover, with the distribution of these chords in the shadow. Chords in both hemispheres are required to achieve the best results. In our case, the higher correlation was 0.81 between $g_c$ and $\epsilon'$, which is expected because we only had chords in the southern hemisphere of the shadow.

We can compare our results with the values determined by Galileo mission images \citep{Nimmo2007}. Our apparent semi-major axis ($a'$ = 1562.0 $\pm$ 3.6 km) was between the axis $a$ (1562.6 km) and $b$ (1560.3 km) of the fitted ellipsoid. Our apparent semi-minor axis ($b'$ = 1560.4 $\pm$ 5.7 km) is nearly the $c$ axis of the ellipsoid (1559.5 km). The expected oblateness was between 0.0019 and 0.0005, in comparison with 0.0010 $\pm$ 0.0028 that we obtained. The equivalent radius ($R_{eq}=a'\sqrt{1 - \epsilon'} = 1561.2 \pm 3.6$ km) agrees with the nominal equivalent radius of 1560.8 $\pm$ 0.3 km of Europa.  According to \cite{Kay2019}, the topographic features of Europa are at a level of some hundreds of metres, therefore it is not possible to infer any topography from our observations.

With the limb fitting, we also obtained a geocentric position for the Europa centre on 2017 March 31$^{}$  at 06:44:00 UTC of

\begin{eqnarray}
\alpha_{J2} &=&  \phantom{-}13^h~12'~15''.548372~\pm~0.96~mas, \nonumber\\
\delta_{J2} &=& -05^{\circ}~56'~48''.687034~\pm~0.62~mas. \label{Eq:positions}
\end{eqnarray}

This position has an offset with respect to the JPL \textit{jup310} ephemerides of $+0.28~mas$ and $+3.81~mas$ for $\Delta\alpha\cos\delta$ and $\Delta\delta,$ respectively, while the offset with respect to the IMCCE NOE-5-2010-GAL \citep{Lainey2009} is $-2.59~mas$ and $-6.01~mas$, respectively, both in the sense of \textit{\textup{observation minus ephemeris}}. More details can be found in Appendix \ref{App:ephem}.



\section{Future Events}\label{Sec:future}

Observing very many occultations by the Galilean satellites is required for better understanding their 3D shape (oblatness and pole positions). This in turn enables obtaining a highly accurate absolute position (with the help of GAIA DR2 catalogue, which furnishes the star positions) that are to be used in dynamical models and with which high-quality ephemerides can be obtained.

We conducted a search for occultation events that would occur between 2017 and 2021, using the GDR2 catalogue and the JPL \textit{jup310} and DE435 ephemerides. In Table \ref{tb:next_ev} we summarise the occultation predictions between 2019 and 2021. The occultation presented here is highlighted. We also list some parameters that are necessary to produce occultation maps such as the closest approach instant (UTC) of the prediction; C/A, the apparent geocentric distance between the satellite and the star at the moment of the geocentric closest approach, in arcseconds; P/A, the satellite position angle regarding the star to be occulted at C/A, in degrees (zero at north of the star, increasing clockwise); magnitude G of the occulted GDR2 star; and finally, the expected magnitude drop ($\Delta mag$) for each event. For more information about the definition and use of these stellar occultation geometric elements, see \cite{Assafin2010,Assafin2012}. We present in Table \ref{tb:next_ev} three events by Io, four by Europa, one by Ganymede, and three by Callisto with $\Delta mag$ higher than 0.005 magnitude. The prediction maps are shown in the Appendix \ref{App:maps}. 

\begin{table}[h]
\begin{center}
\caption{Predicted stellar occultations by the Galilean moons between 2019 and 2021.}
\begin{tabular}{lccrcr}
\hline
\hline
Sat. & Time (UTC) &  C/A &  P/A &  $Gmag^{\star}$  &  $\Delta mag$
 \\
\hline
\rowcolor{lightgray}
502   & 2017-03-31 06:44 & 0.09 & 19.98  & \phantom{0}9.5 & 0.030\\
502   & 2019-05-06 20:32 & 0.58 & 183.07 & 10.9 & 0.008\\
502   & 2019-06-04 02:26 & 0.12 &   4.46 & \phantom{0}9.1 & 0.037\\
504   & 2019-06-05 23:12 & 0.90 & 182.86 & 10.2 & 0.020\\
501   & 2019-09-09 02:33 & 0.26 & 189.87 & 11.0 & 0.008\\
501   & 2019-09-21 13:08 & 1.22 &   7.97 & 11.3 & 0.007\\
504   & 2020-06-20 16:03 & 1.45 & 348.44 & 10.9 & 0.012\\
502   & 2020-06-22 02:09 & 2.07 & 348.23 & 11.3 & 0.005\\
501   & 2021-04-02 10:24 & 1.02 & 344.64 & \phantom{0}5.8 & 0.740\\
503   & 2021-04-25 07:55 & 0.44 & 160.79 & 11.1 & 0.005\\
504   & 2021-05-04 23:01 & 0.90 & 342.43 & 10.4 & 0.027\\
\hline
\hline
\multicolumn{6}{l}{\textbf{Note}: See text for the definitions of geometric elements.}
\label{tb:next_ev}
\end{tabular}
\end{center}
\end{table}

\begin{figure}
\includegraphics[width=0.5\textwidth]{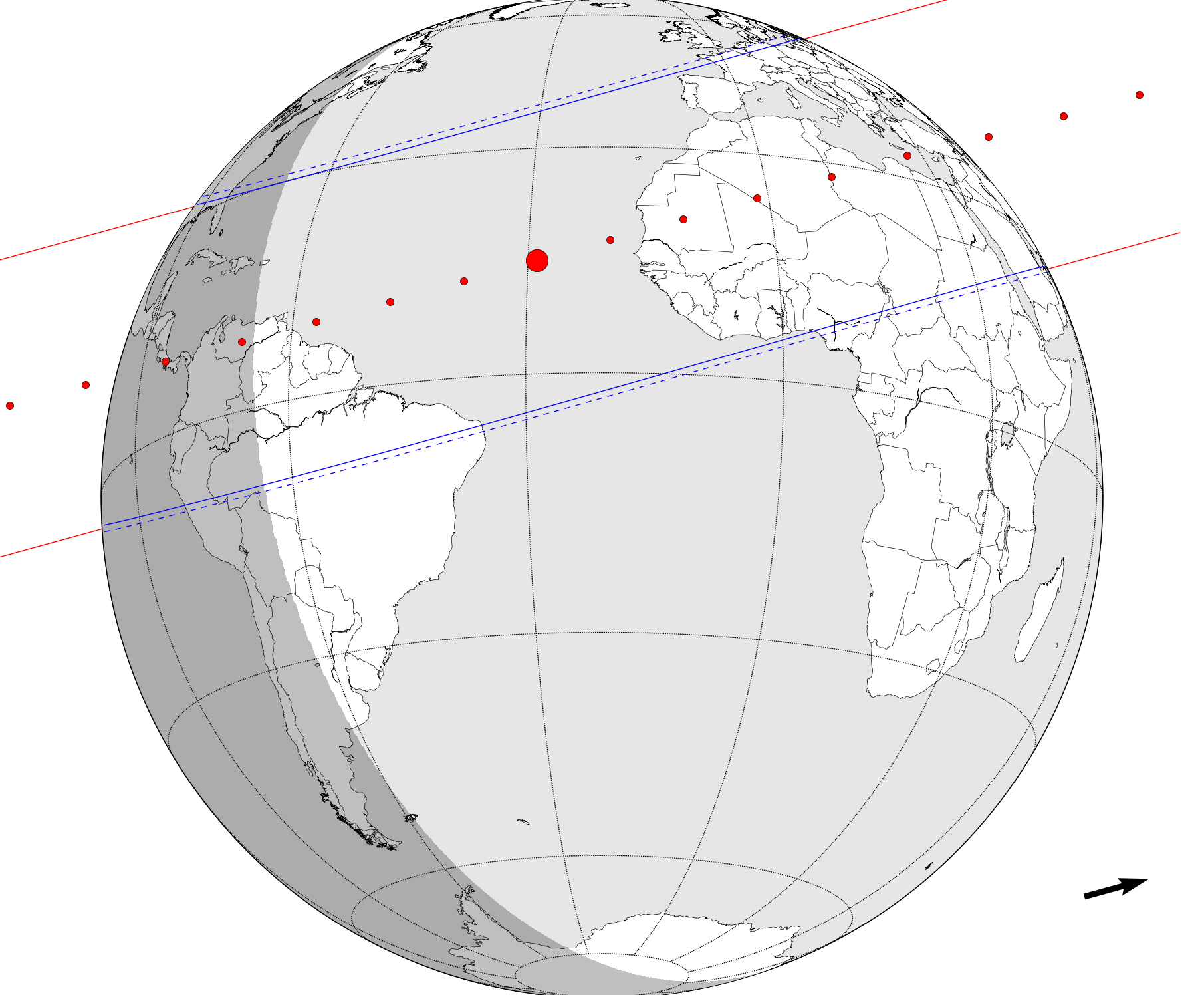}
\caption{Prediction of the occultation by Io on 02 April 2021 10:24 UTC, using JPL DE435 and \textit{jup310} ephemeris. The blue lines represent the expected size limit of Europa. The blue dashed lines are an uncertainty of 20 mas in Europa's position. The red dots represents the centre of the body for a given time, each separated by 1 minute. The arrow represents the sense of the shadow's motion. The dark grey and the light grey area represent the night and the day part of the globe, respectively.}
\label{Fig:Io}
\end{figure}

We draw the attention to the occultation by Io of a mag. G 5.8 star on 2012 April 2. The magnitude drop will be about 0.740 magnitude. This event can be observed in the South of North America, Central America, and in the North of South America (Fig. \ref{Fig:Io}).

This type of observation is challenging because of the low magnitude drop. We suggest the use of red filters (or a narrow-band methane filter, centred on 889 nm with a width of 15 nm) in these observations. Jupiter should be placed outside the field of view of the camera when possible, which will reduce noise effects caused by scattered light from Jupiter. It is very important to test the equipment configuration (gain, exposure time, etc.) some days before the occultation in order to ensure the best S/N possible.

We also suggest the use of fast cameras with high-cadence images and minimum readout time. Stacking images techniques, such as the one applied here (detailed in Appendix \ref{App:stack}), can be used to increase the S/N of the images. Telescopes as small as 20 cm (8'') can provide useful data. Adaptive optics is also a possibility, with adequate exposure time. For a telescope with a large aperture and a sensitive camera, this could result in the direct detection of a thin atmosphere around a Galilean satellite. Observations with resolutions higher than the Fresnel scale (about 0.44 km at Jupiter) can result in the direct measurement of refraction spikes by a thin atmosphere such as the spikes observed in occultations by Pluto \citep{Dias-Oliveira2015,Sicardy2016,Meza2019}.

\section{Discussion}\label{Sec:conc}

Stellar occultation is a ground-based technique capable of determining positions, sizes, and shapes with uncertainties of some kilometres. These values are comparable with data obtained with space probes. In this work we took advantage of the beginning of the Jupiter passage through a very dense stellar region, with the Galactic centre as background, and organised an observational campaign to obtain data from a predicted stellar occultation by the Galilean moon Europa. Its ephemeris indicates that it would pass in front of a mag. G = 9.5 GDR2 star on 2017 March 31. Out of the nine stations in eight sites across South America, three obtained a positive detection of the event.

The fitted ellipse for Europa gives an area equivalent radius of 1561.2 $\pm$ 3.6 km. Its absolute position uncertainty is 0.80 mas (2.55 km), representing an offset with respect to the JPL \textit{jup310} ephemeris of 3.82 mas and of 6.54 mas with respect to the NOE-5-2010-GAL ephemeris. This position for Europa at the time of the occultation took advantage of the high-precision position of Gaia DR2 stars and the fact that stellar occultations render accurate  body-star relative positions.

This is the first reported observation of a stellar occultation by Europa. The favourable configuration of Jupiter, which has the Galactic plane as background, increases the chances of observing other bright star occultations by its main satellites, as discussed in Sect. \ref{Sec:future}. This configuration will only occur again in 2031. For the future events predicted in the 2019 and 2021, campaigns will be organised in due time. 

We encourage the observational community to observe the predicted future stellar occultations. Even the amateur community with telescopes as small as 20 cm (8'') can help provide useful data. A stellar occultation provides an apparent size and shape at a specific moment. Combined with the observations of the upcoming events, a 3D shape of these moons can be acquired with kilometre precision and complement space mission data. This information can aid in the study of planetary formation, evolution, and the influence of tides raised by the moons' primary in their orbit. This highlights the importance of these events.

Finally, all this information can also be used together with dynamical models to ensure highly accurate orbits for these moons. These orbits can be helpful for future space probes aimed at the Jovian system, such as JUICE and the Europa Clipper mission.

\begin{acknowledgements}
    This study was financed by the Coordena\c{c}\~ao de Aperfei\c{c}oamento de Pessoal de N\'ivel Superior - Brasil (CAPES) - Finance Code 001. Part of this research is suported by INCT do e-Universo, Brazil (CNPQ grants 465376/2014-2). Based in part on observations made at the Laborat\'orio Nacional de Astrof\'isica (LNA), Itajub\'a-MG, Brazil. BM thanks the CAPES/Cofecub-394/2016-05 grant. G.B.R. acknowledges the support of the CAPES and FAPERJ/PAPDRJ (E26/203.173/2016) grants. ARGJ thanks FAPESP proc. 2018/11239-8. MA thanks CNPq (Grants 427700/2018-3, 310683/2017-3 and 473002/2013-2) and FAPERJ (Grant E-26/111.488/2013). VL's research was supported by an appointment to the NASA Postdoctoral Program at the NASA Jet Propulsion Laboratory, California Institute of Technology, administered by Universities Space Research Association under contract with NASA. RVM acknowledges the grants CNPq-306885/2013, CAPES/Cofecub-2506/2015, FAPERJ/PAPDRJ-45/2013 and  FAPERJ/CNE/05-2015. JIBC acknowledges CNPq grant 308150/2016-3. FBR acknowledges CNPq support, proc. 309578/2017-5. RS e OCW acknowledges Fapesp proc. 2016/24561-0, CNPq proc. 312813/2013-9 and 305737/2015-5.

\end{acknowledgements}

\bibliographystyle{aa} 
\bibliography{ref} 

\begin{appendix} 
\section{Stacking image procedure}\label{App:stack}

The stacking image procedure was developed with the aim to attenuate the image noise, thereby increasing the image S/N when is not possible to just integrate over more time. This is the case when the Galilean moons are observed without a filter, as the brightness of Jupiter or even the satellite itself would quickly saturate the CCD image before we achieve an adequate S/N.

The first step in this procedure is to chose a well-sampled object in the images that is to be used as reference to align the images for the stacking. This object should not have a significant motion (in the sky plane) between the first and last image. In our case, in the absence of stars, we used Europa to align the images. Each individual image had an exposure of 0.05 seconds and the same cycle, and the motion of Europa was 5.39 mas/s in the sky plane, corresponding to a motion of 0.0074 pixels/s on the CCD.   

We measured the object centroid ($x,y$) with a 2D circular symmetric Gaussian fit over pixels within one full-width at half-maximum (FWHM $\propto$ seeing) from the centre. This was done using the \textsc{PRAIA} package \citep{Assafin2011}. The uncertainty in the centroid measurement was about 36 mas (0.05 pixels). The alignment consists of vertical and horizontal shifts for each image ($\Delta x$,$\Delta y$) relative to a chosen reference image; in our case, the first image.

\begin{figure}
\centering
\includegraphics[width=0.50\textwidth]{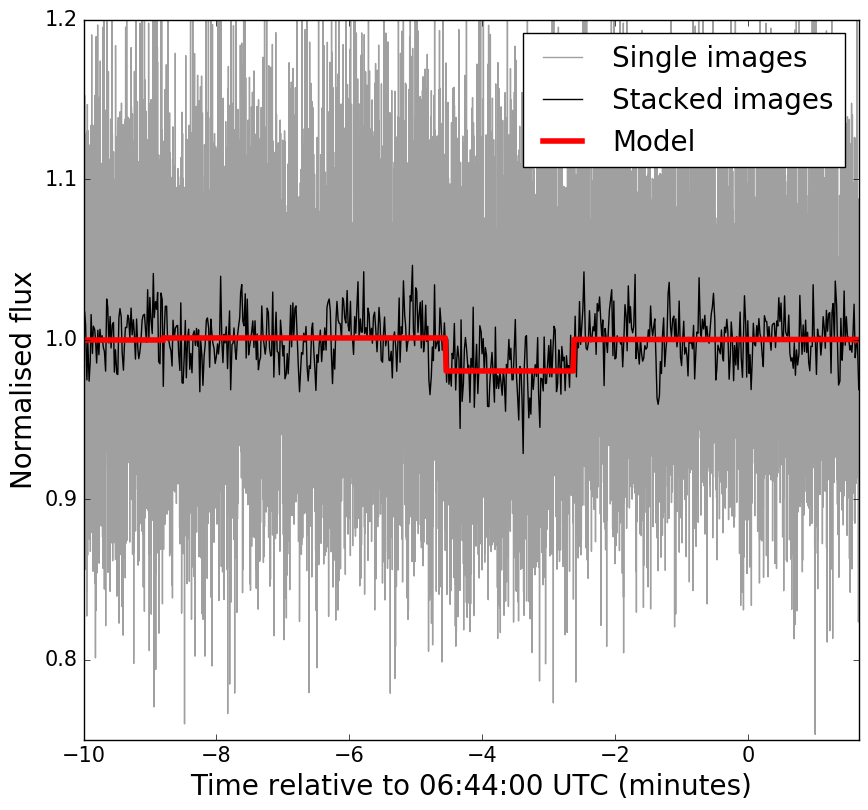}
\caption{Normalised light flux from FOZ images with (black line) and without (light gray line) the stacking image procedure. The red line is the modelled light curve. Each stacked image contains the data of 20 single images stacked together.}
\label{Fig:stk_phot}
\end{figure}

We performed a stack procedure over each of the 20 images, which resulted in a temporal resolution of 1.0 second. Fig. \ref{Fig:stk_phot} contains the light curve we obtained using all individual frames (light gray line), the light curve using the stacked images (black line), and the modelled light curve after the fit (red line). Before our procedure, the normalised light curve rms was 0.073, while after the stacking, we were able to achieve 0.016. It is important to highlight that the expected drop would be about 0.021 in flux. Without the stacking image procedure, it would be impossible to determine the ingress and/or the egress times.

\section{Light-curve fitting procedure}\label{App:fitting}

The most common geological model of Europa is an ice surface on top of a global ocean \citep{Anderson1998}. The presence of plumes of water vapour \citep{Roth2014,Jia2018,Sparks2016,Sparks2019} is a strong evidence of this model.

The OPD/B\&C light curve presents an apparent smooth flux decay on the ingress and egress times during the passage of the body in front of the star, and this slope would be compatible with the common picture of the Europa geology and water vapour evidence. These features must be analysed with caution, however, because the measurements show some level of correlated noise. To objectively assess the significance of this slope, we tested two scenarios: one simple (sharp-edge model) scenario in which we assumed that there are no slopes during the ingress and egress times, and one more complex scenario (trapezoidal model) with a smooth ingress and egress. It is also straightforward to show that the sharp-edge model is a particular case of a trapezoidal model, allowing us to analyse the decrease in residuals when the number of fitted parameters is increased.

Because the more complex model has fewer degrees of freedom, it should at least fit the observables as well as the more straightforward case. Having set the characteristics of the problem, we used the Fisher-Snedecor F-test \citep{Chow1960,Seber2003} to estimate the probability that the trapezoidal model fits the data better than a more simple sharp-edge model. Under the null hypothesis, the trapezoidal case does not produce a better fit than the standard case, or it could be argued that the slope coefficients are similar. The test follows an F-distribution, and the null hypothesis is rejected when the calculated F value is higher than the expected critical value (or significance).

As an example, we discuss the light curve obtained by OPD/B\&C. The normalised chi-square statistics obtained from the sharp edge model was 1.06, whereas the value obtained with the trapezoidal model was 1.04. These models were applied to the data within an interval of five minutes centred at the central instant for each station. Using the F distribution, we can predict that within a 95\% confidence level, the computed chi-square value for the plume model should lie around 0.74. In this case, increasing the number of model parameters leads only to a better fit of the noise because we found out that the probability of the trapezoidal model to explain the data better is about 17\%. Our measurements therefore do not support any further information about the detection of a thin atmosphere (or a plume). The FOZ and OBSPA light curves have lower cadences and therefore a lower time resolution during ingress and egress of the event. The probabilities for these sites are even lower, about 1\% and 3\% respectively, as expected.

\begin{figure}
\centering
\includegraphics[width=0.50\textwidth]{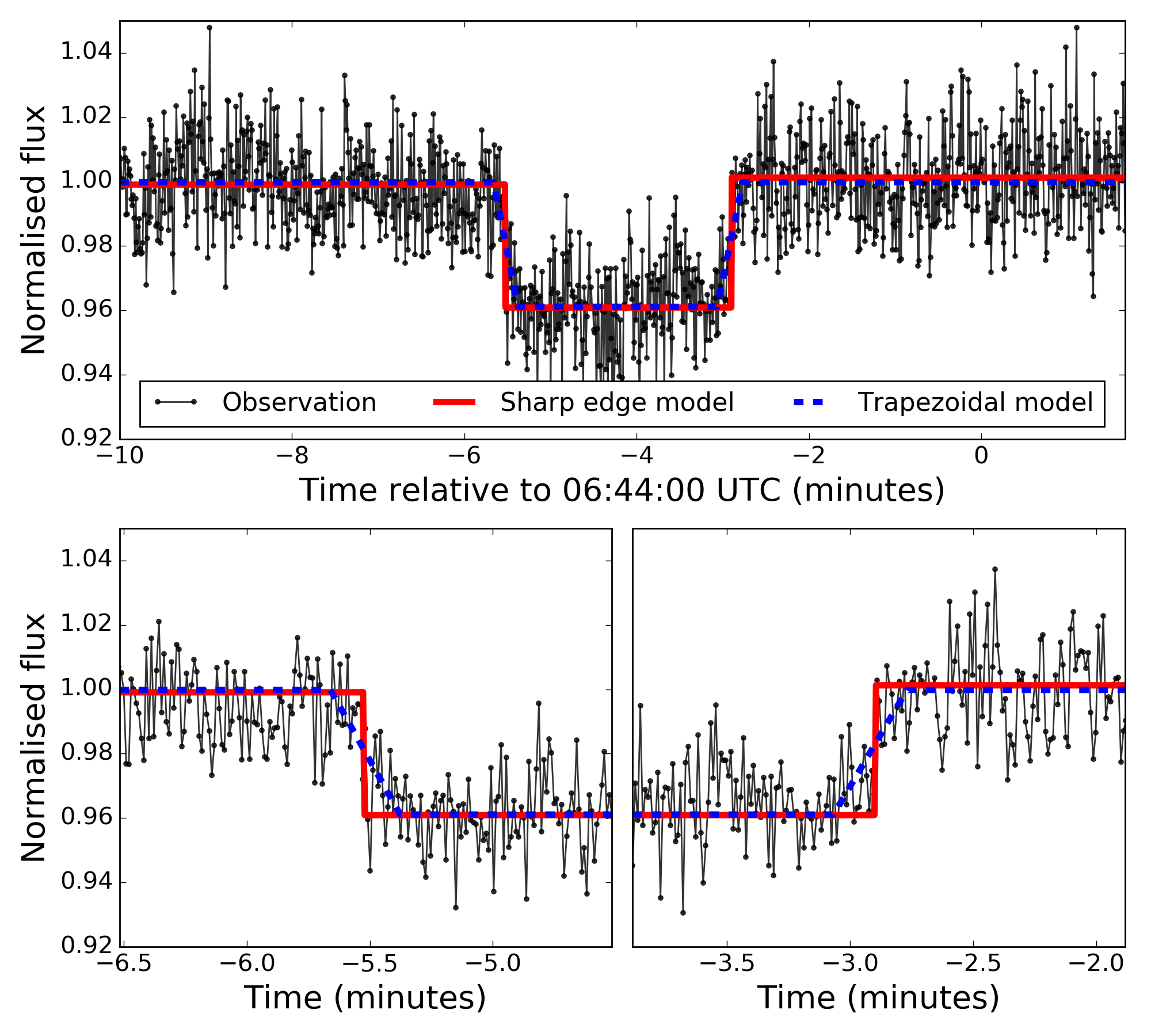}
\caption{Normalised light curve obtained at OPD/B\&C with a zoom-in at the moment of the occultation. The black dots show the data, while the red lines represent the sharp-edge model for each observation. Tthe dotted blue line shows the trapezoidal model.}
\label{Fig:zoom_opd}
\end{figure}

\begin{figure}
\centering
\includegraphics[width=0.50\textwidth]{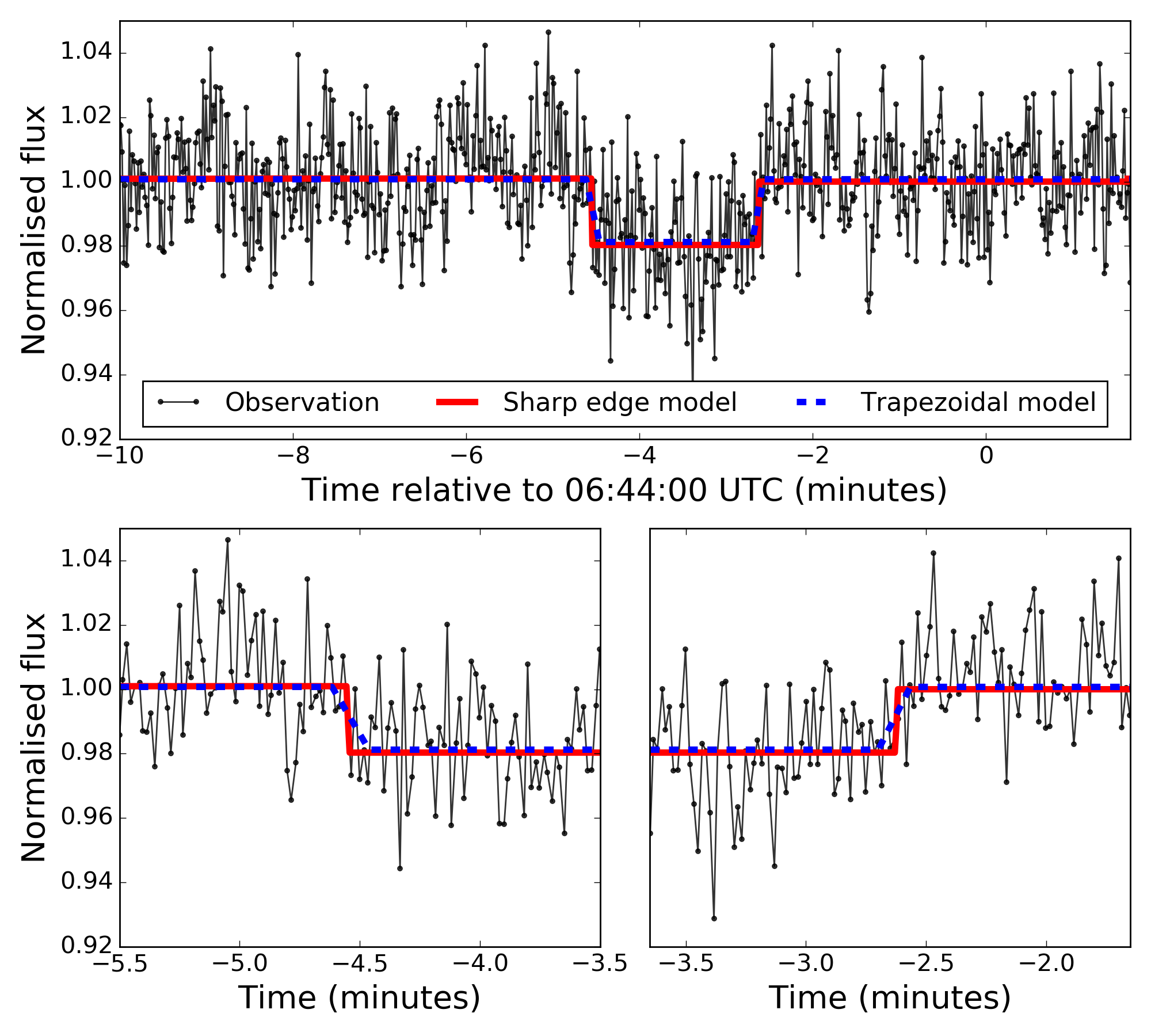}
\caption{Same as Fig.~\ref{Fig:zoom_opd} for the light curve obtained at FOZ.}
\label{Fig:zoom_foz}
\end{figure}

\begin{figure}
\centering
\includegraphics[width=0.50\textwidth]{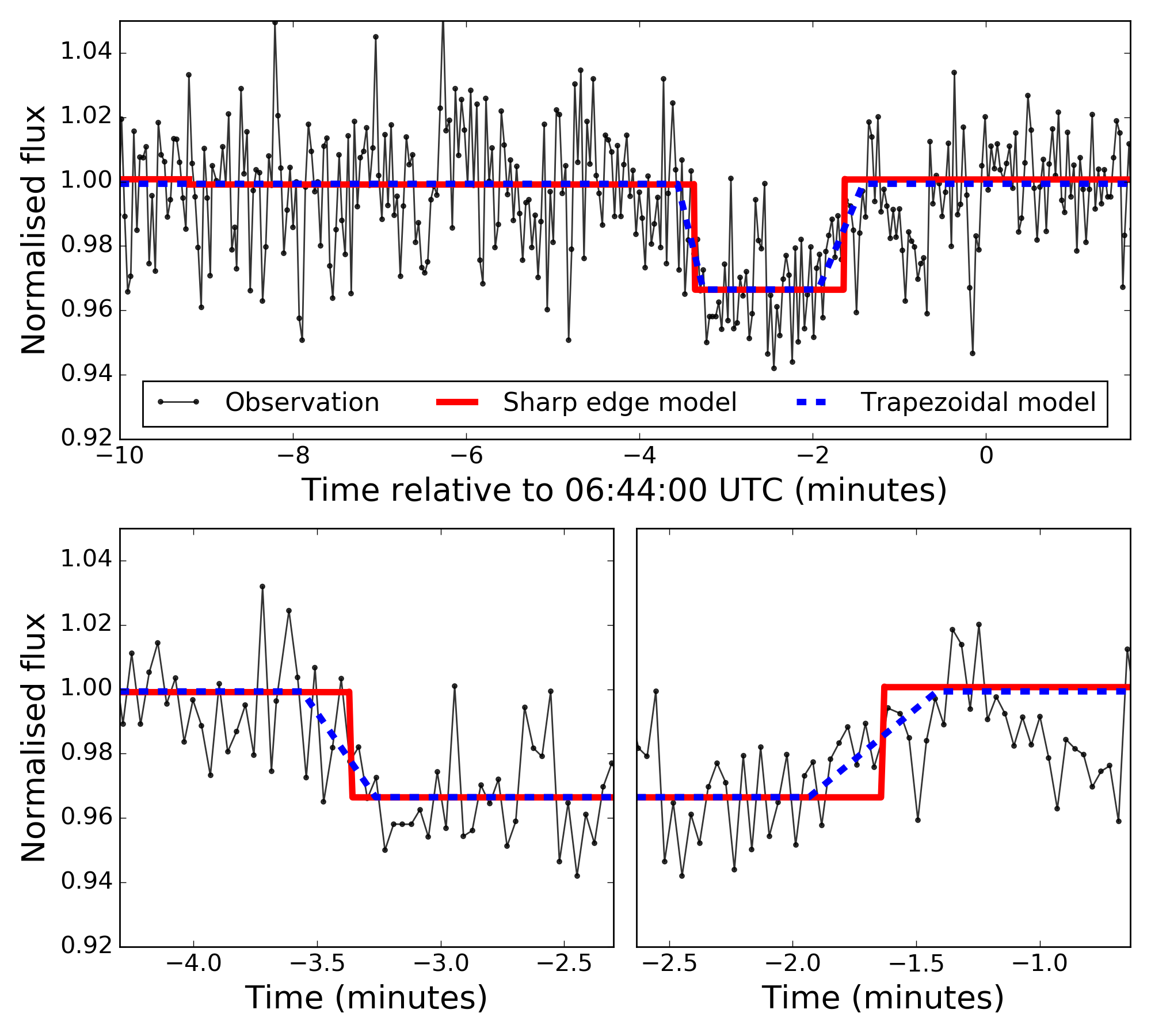}
\caption{Same as Fig.~\ref{Fig:zoom_opd} for the light curve obtained at OBSPA.}
\label{Fig:zoom_spa}
\end{figure}

Figures \ref{Fig:zoom_opd}, \ref{Fig:zoom_foz}, and \ref{Fig:zoom_spa} contain the individual light curves for each observational site (OPD/B\&C, FOZ, and OBSPA, respectively) with a zoom (bottom graphs) of two minutes centred on the ingress and egress times. The x-axes are the time in minutes relative to 06:44:00 UTC, and the y-axes are the normalised light flux. The red lines are the sharp-edge model for each observation, and the dotted blue lines are the trapezoidal model.

\section{Comparison between ephemeris} \label{App:ephem}

The absolute position of Europa obtained here has an uncertainty smaller than 1 mas ($\sim$ 3 km). This level of accuracy is of the same order of magnitude as those obtained from space mission images \citep{Tajeddine2015}. However, our position is completely independent of any variation in the satellite albedo and is not affected by phase angle effects. Our geocentric position for Europa is given in Eq. \eqref{Eq:positions_apendix},  

\begin{eqnarray}
\alpha_{J2} &=&  \phantom{-}13^h~12'~15''.548372~\pm~0.96~mas, \nonumber\\
\delta_{J2} &=& -05^{\circ}~56'~48''.687034~\pm~0.62~mas. \label{Eq:C1} \label{Eq:positions_apendix}
\end{eqnarray}

Here we compare this position ($\alpha_{J2}, \delta_{J2}$) with different geocentric ephemeris ($\alpha_{ephem}, \delta_{ephem}$) using Eq. \eqref{Eq:offset_apendix}, 

\begin{eqnarray}
\Delta\alpha &=& \alpha_{J2} - \alpha_{ephem}~, \nonumber\\
\Delta\delta &=& \delta_{J2} - \delta_{ephem}~, \nonumber\\
s &\simeq& \sqrt{\Delta\alpha^2\cos^2 \left(\dfrac{\delta_{J2} + \delta_{ephem}}{2}\right) + \Delta\delta^2} \label{Eq:offset_apendix}~.
\end{eqnarray}

In Table \ref{tb:ephem} we present the offset between the Europa position as derived based on this stellar occultation and the ephemeris. We compare our result with 12 ephemerides combinations. We use 6 planetary ephemerides (from the JPL, DE438s, DE436s, DE435, and DE430; and from the IMCCE, INPOP17a and INPOP13c) and two satellite ephemerides (from the JPL,  \textit{jup310}; and from the IMCCE, NOE-5-2010-GAL). 

\begin{table}[h]
\begin{center}
\caption{Offset between the geocentric position of Europa obtained from the stellar occultation on 2017 March 31$^{}$ at 06:44 UTC and different ephemerides in the sense \textit{\textup{observation minus ephemeris}}.}
\begin{tabular}{cccc}
\hline
\hline
 & & JPL \textit{jup310} & IMCCE NOE-5-2010-GAL  \\
 & & (mas) & (mas) \\
\hline
\hline
         & $\Delta\alpha$ & \phantom{+}-\phantom{0}7.94 & \phantom{+}-\phantom{0}5.07 \\
DE438s   & $\Delta\delta$ & \phantom{+}-10.91 & \phantom{+}-\phantom{0}1.08 \\
         & $s$            &  \phantom{-+}13.46 &  \phantom{+-0}5.15 \\
\hline
         & $\Delta\alpha$ & \phantom{+}-\phantom{0}6.68 & \phantom{+}-\phantom{0}3.80 \\
DE436s   & $\Delta\delta$ & \phantom{+}-\phantom{0}7.59 & \phantom{-}+\phantom{0}2.23 \\
         & $s$            &  \phantom{-+}10.09 &  \phantom{+-0}4.39 \\
\hline
         & $\Delta\alpha$ & \phantom{+}-\phantom{0}0.27 & \phantom{-}+\phantom{0}2.60 \\
DE435    & $\Delta\delta$ & \phantom{+}-\phantom{0}3.81 & \phantom{-}+\phantom{0}6.01 \\
         & $s$            &  \phantom{-+0}3.82 &  \phantom{+-0}6.54 \\
\hline
         & $\Delta\alpha$ & \phantom{+}-\phantom{0}1.67 & \phantom{-}+\phantom{0}1.19 \\
DE430    & $\Delta\delta$ & \phantom{+}-\phantom{0}3.69 & \phantom{-}+\phantom{0}6.13 \\
         & $s$            &  \phantom{+-0}4.06 &  \phantom{+-0}6.24 \\
\hline
         & $\Delta\alpha$ & \phantom{-}+\phantom{0}0.63 & \phantom{+}-\phantom{0}2.23 \\
INPOP17a & $\Delta\delta$ & \phantom{+}-\phantom{0}8.59 & \phantom{-}+\phantom{0}1.23 \\
         & $s$            &  \phantom{+-0}8.61 &  \phantom{+-0}2.54 \\
\hline
         & $\Delta\alpha$ &  \phantom{+}-\phantom{0}2.64 & \phantom{+}-\phantom{0}5.50 \\
INPOP13c & $\Delta\delta$ & \phantom{+}-10.28 & \phantom{+}-\phantom{0}0.46 \\
         & $s$            &  \phantom{-+}10.61 &  \phantom{-+0}5.49 \\
\hline
\hline
\label{tb:ephem}
\end{tabular}
\end{center}
\end{table}

\section{Future event maps}\label{App:maps}

In this appendix we present the stellar occultation maps for the predicted events that will occur between 2019 and 2021 (See Table \ref{tb:next_ev} for more details). We indicate in each figure the satellite that will occult the star, and the date and time of the occultation in UTC. The blue lines represent the expected size limit of the occulting satellite. The red dots are the centre of the body for a given time, each separated by 1 minute. The large red dot represents the centre of the body at the time of closest approach (CA) indicated in the figure label. The arrow represents the direction of motion of the shadow. The dark grey and light grey areas represent the night and day part of the globe, respectively.


\begin{figure}[h]
\includegraphics[width=0.45\textwidth]{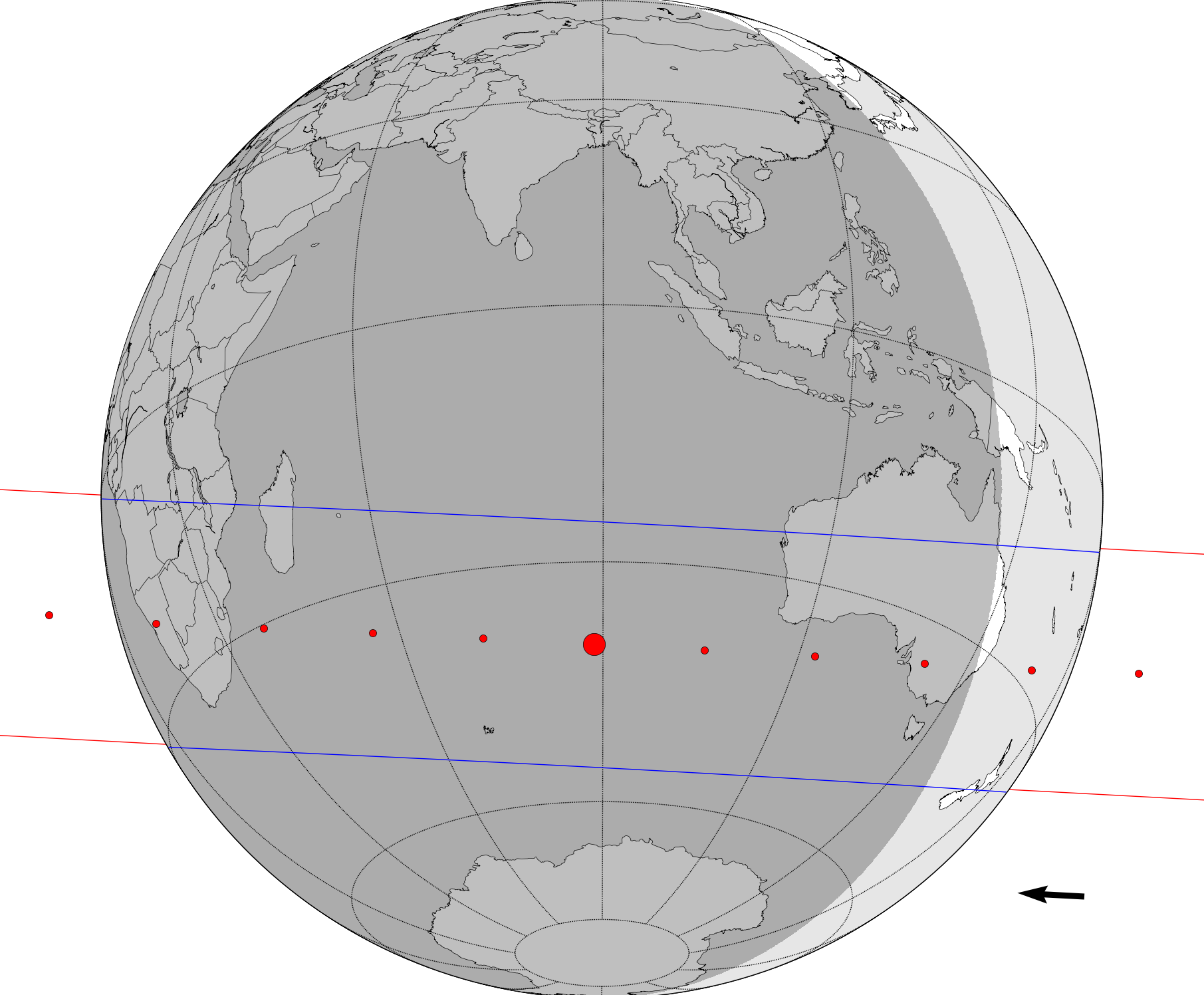}
\caption{Occultation of a mag. G 10.9 star by Europa on 2019 May 5, 20:32 UTC. The predicted relative velocity of the event is 23.1 km/s.}
\label{Fig:E1}
\end{figure}               

\begin{figure}[h]
\includegraphics[width=0.45\textwidth]{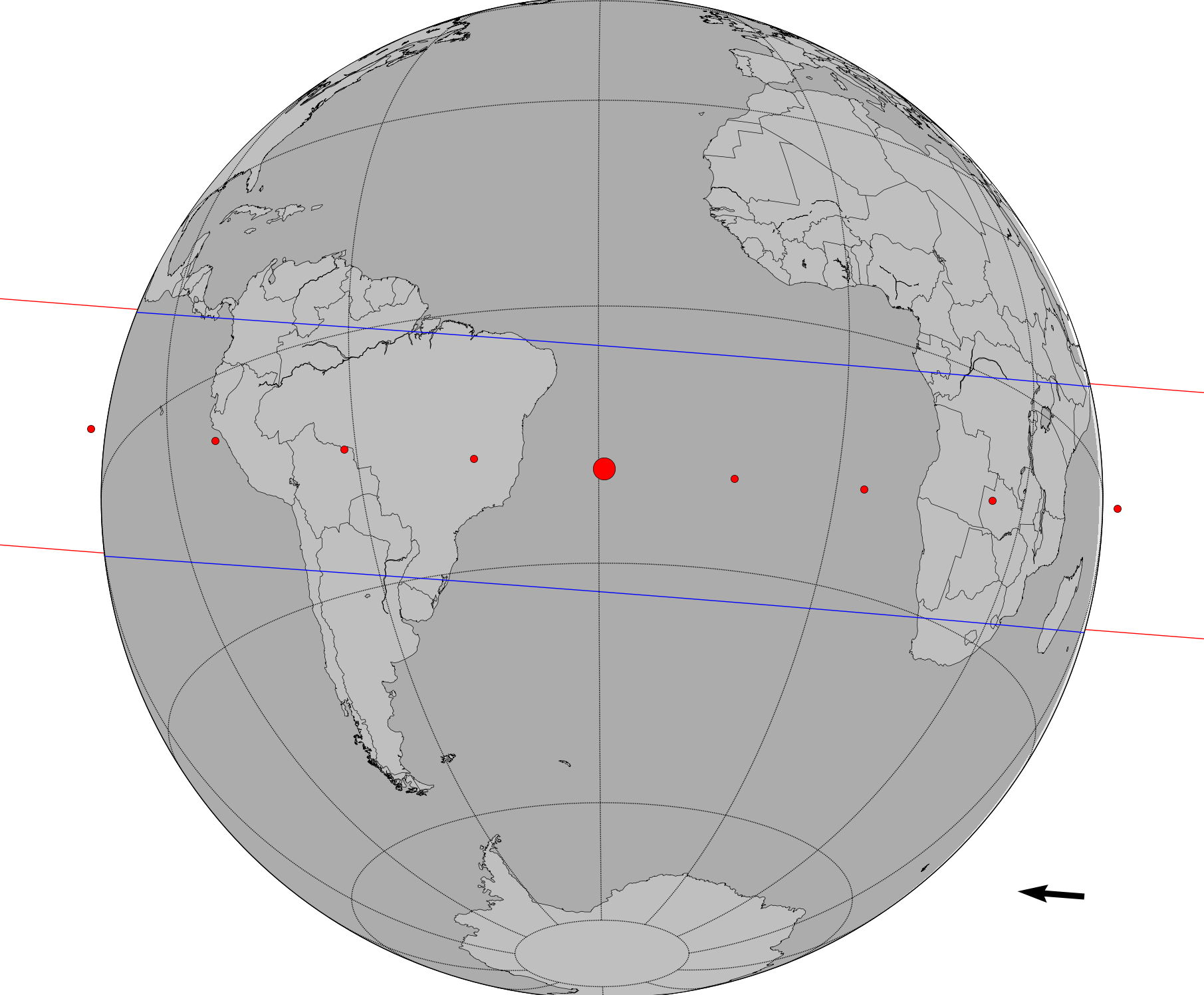}
\caption{Occultation of a mag. G 9.1 star by Europa on 2019 June 4, 02:26 UTC. The predicted relative velocity of the event is 27.3 km/s.}
\label{Fig:E2}
\end{figure}               

\begin{figure}[h]
\includegraphics[width=0.45\textwidth]{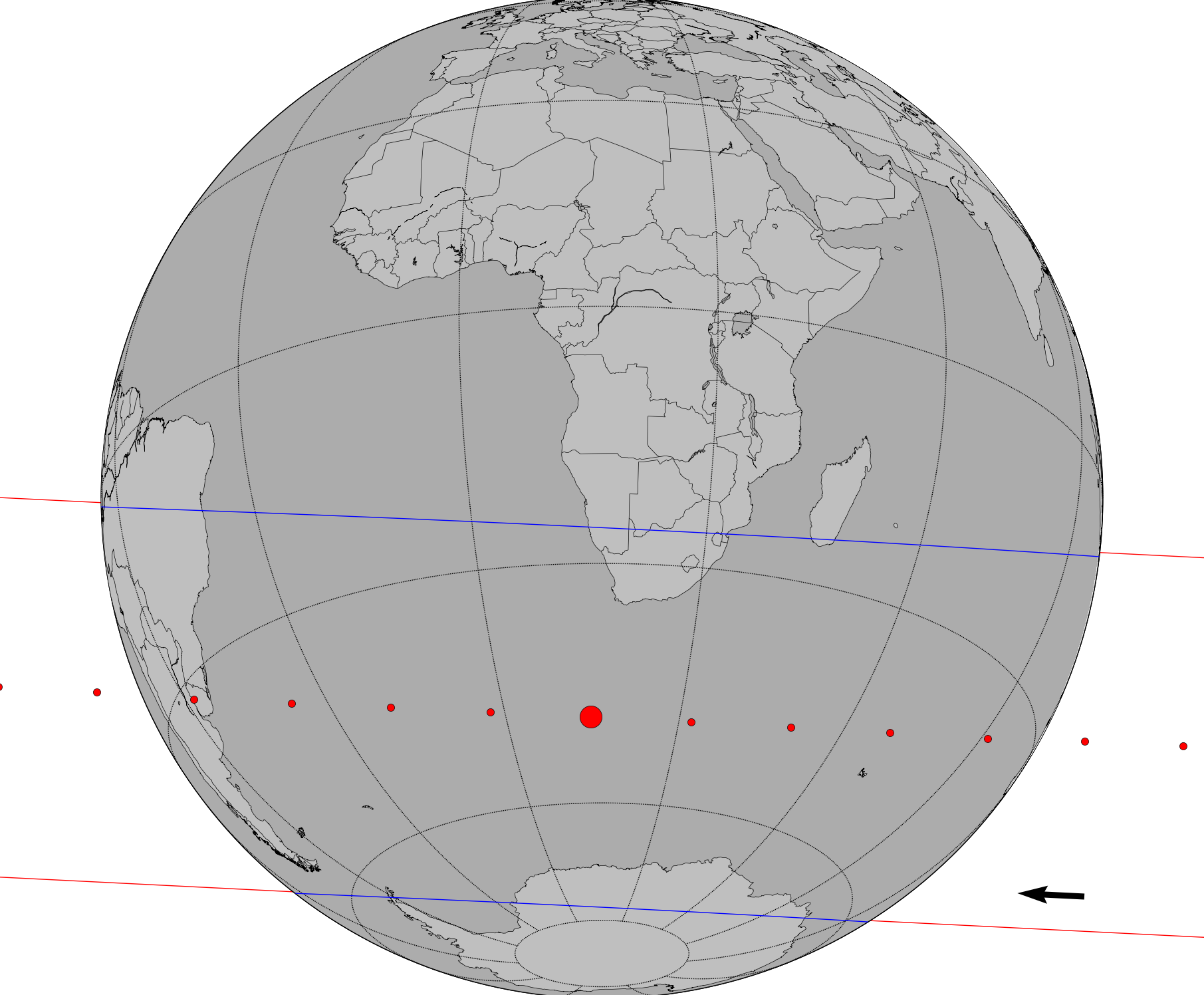}
\caption{Occultation of a mag. G 10.2 star by Callisto on 2019 June 5, 23:12 UTC. The predicted relative velocity of the event is 21.0 km/s.}
\label{Fig:C1}
\end{figure}               

\begin{figure}[h]
\includegraphics[width=0.45\textwidth]{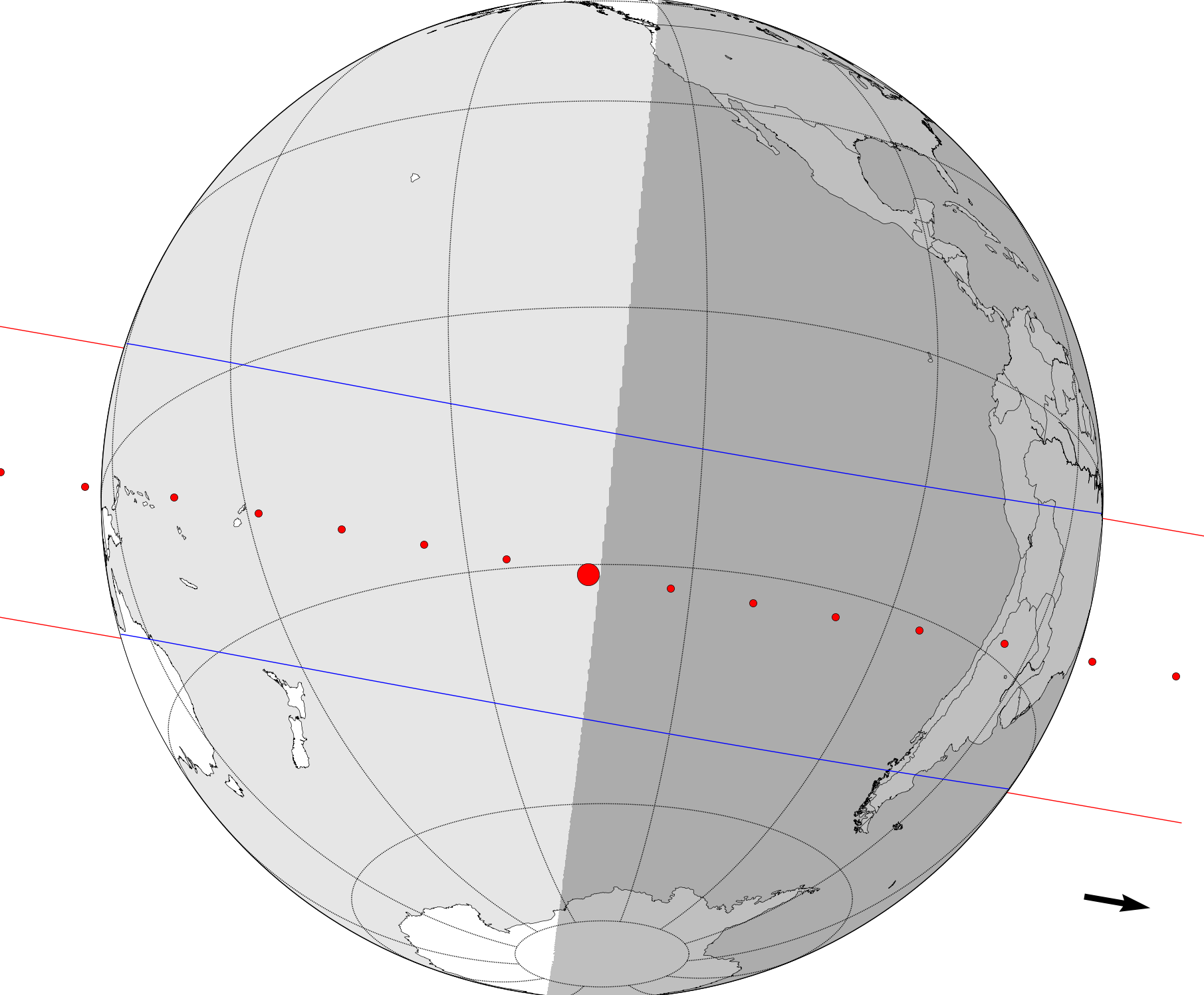}
\caption{Occultation of a mag. G 11.0 star by Io on 2019 September 9, 02:33 UTC. The predicted relative velocity of the event is 18.1 km/s.}
\label{Fig:I1}
\end{figure}               

\begin{figure}[h]
\includegraphics[width=0.45\textwidth]{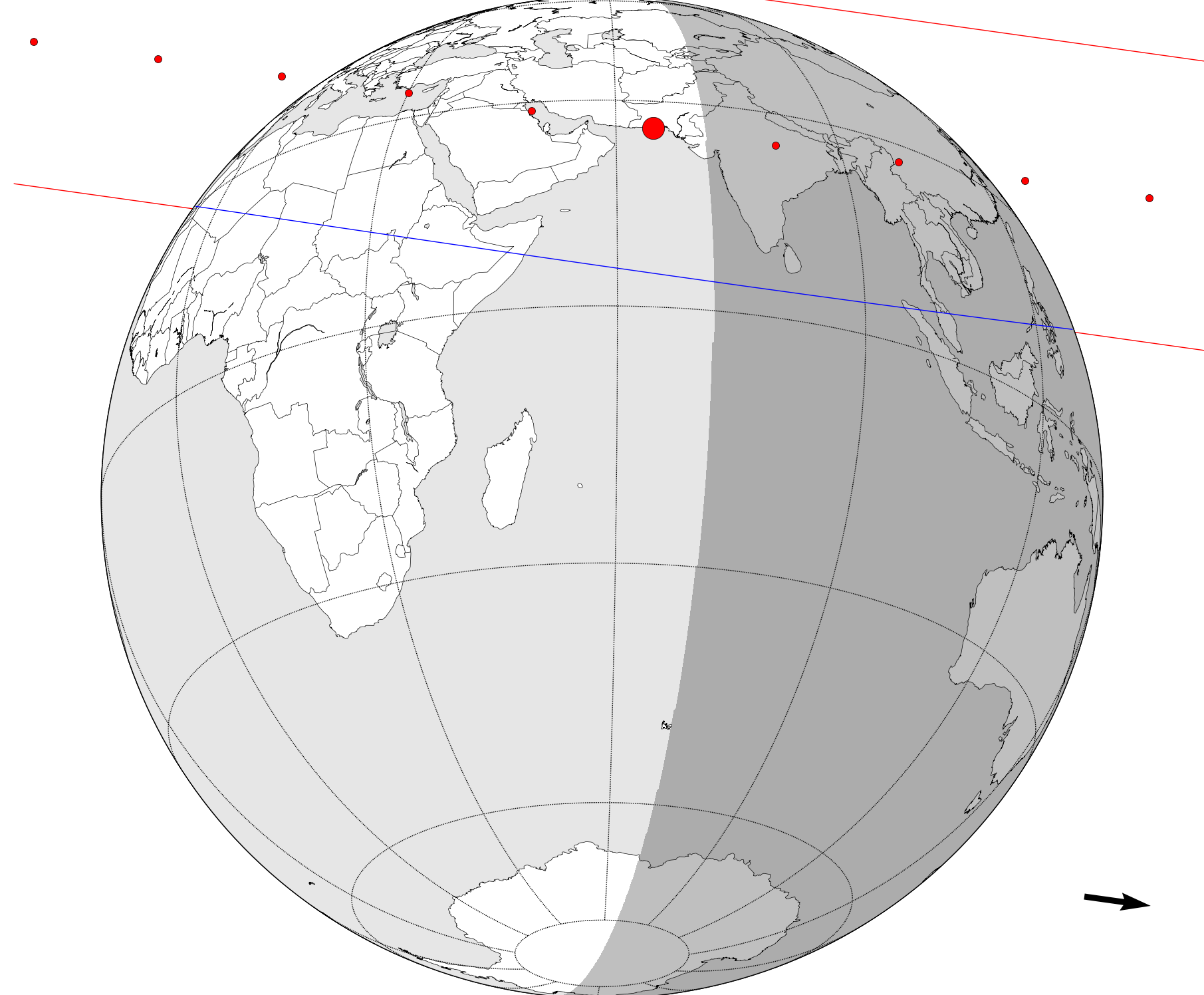}
\caption{Occultation of a mag. G 11.3 star by Io on 2019 September 21, 13:08 UTC. The predicted relative velocity of the event is 26.5 km/s.}
\label{Fig:I2}
\end{figure}


\begin{figure}[h]
\includegraphics[width=0.45\textwidth]{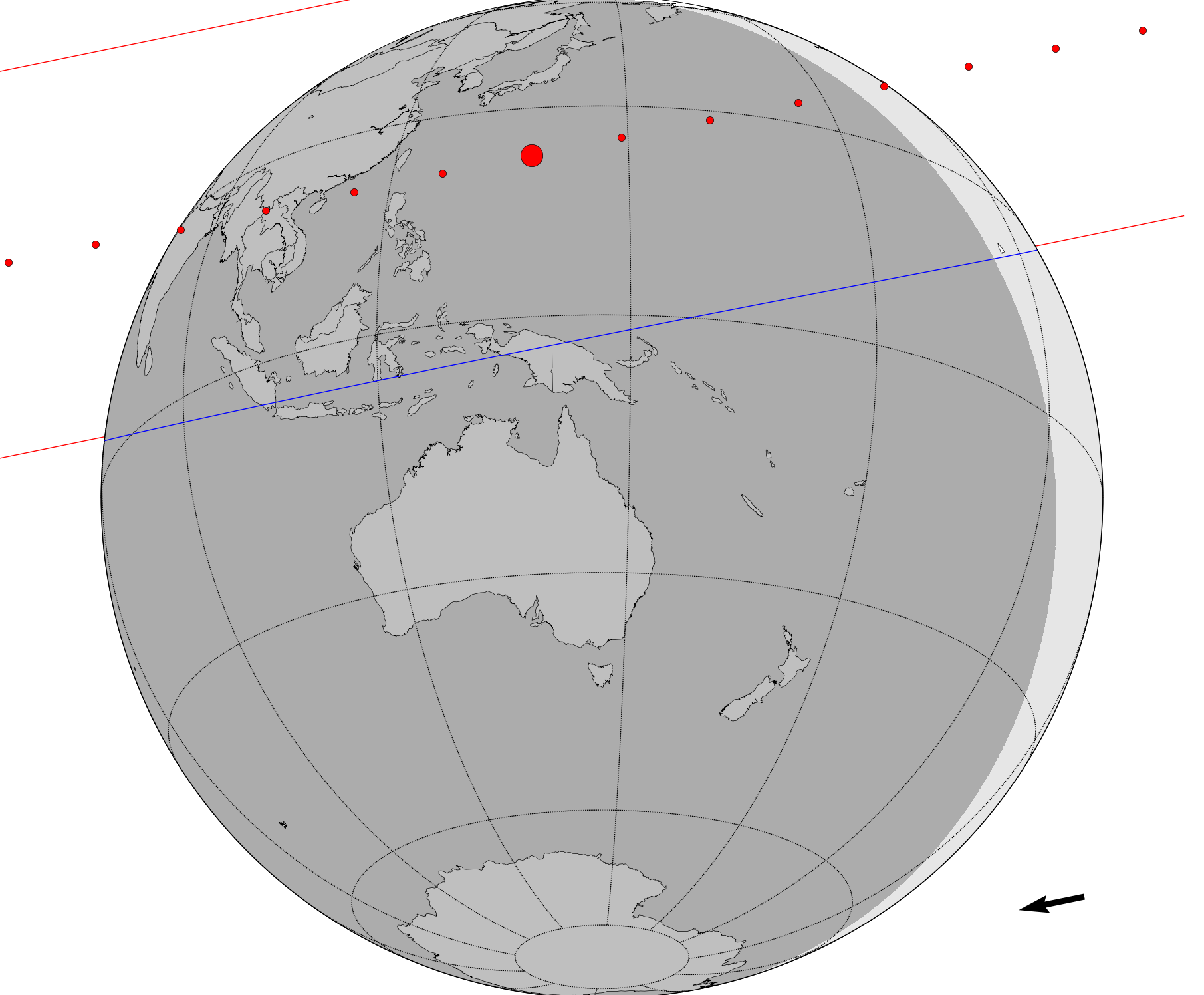}
\caption{Occultation of a mag. G 10.9 star by Callisto on 2020 June 20, 16:03 UTC. The predicted relative velocity of the event is 18.9 km/s.}
\label{Fig:C2}
\end{figure}               

\begin{figure}[h]
\includegraphics[width=0.45\textwidth]{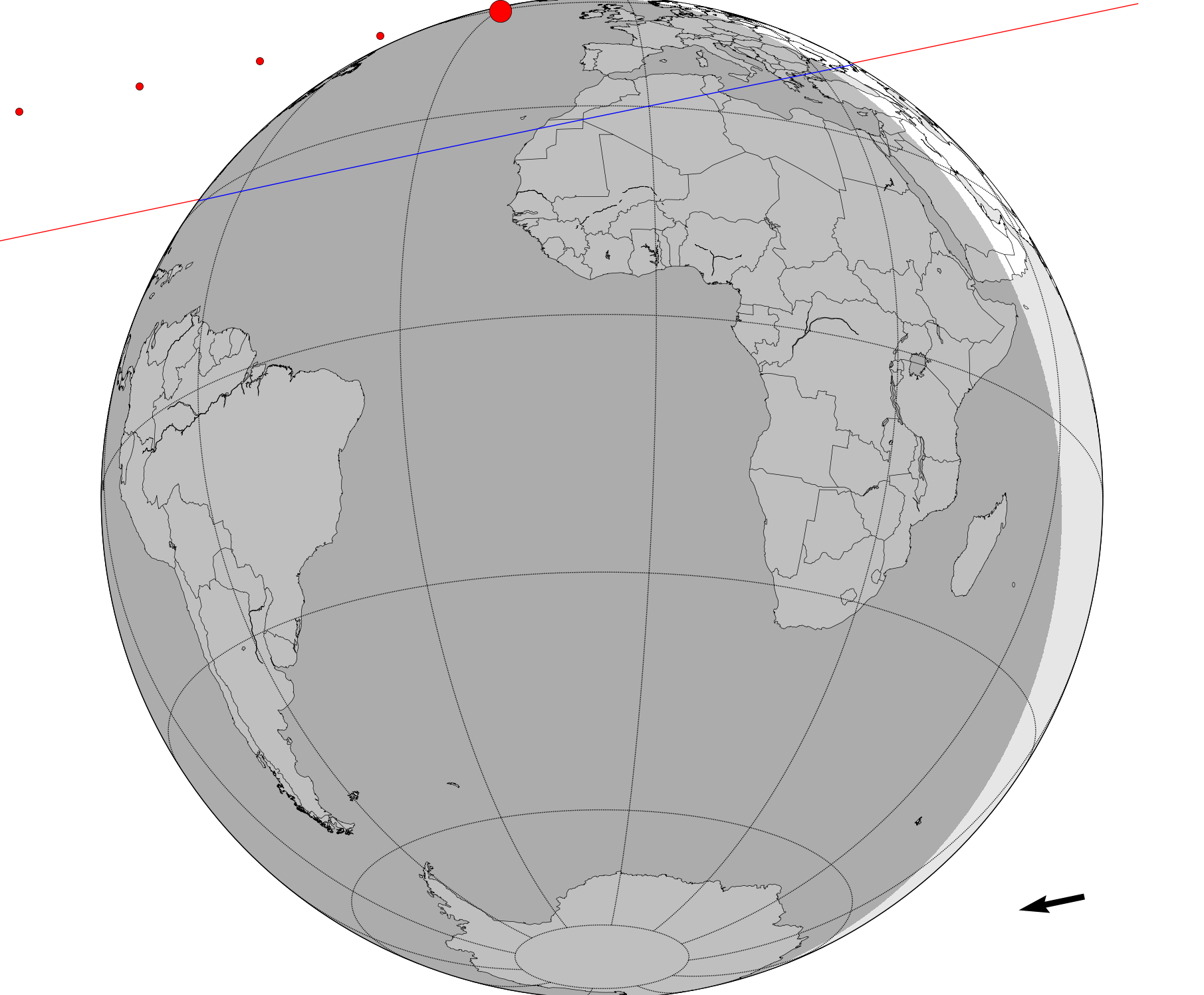}
\caption{Occultation of a mag. G 11.3 star by Europa on 2020 June 22, 02:09 UTC. The predicted relative velocity of the event is 26.1 km/s.}
\label{Fig:E3}
\end{figure}               


\begin{figure}[h]
\includegraphics[width=0.45\textwidth]{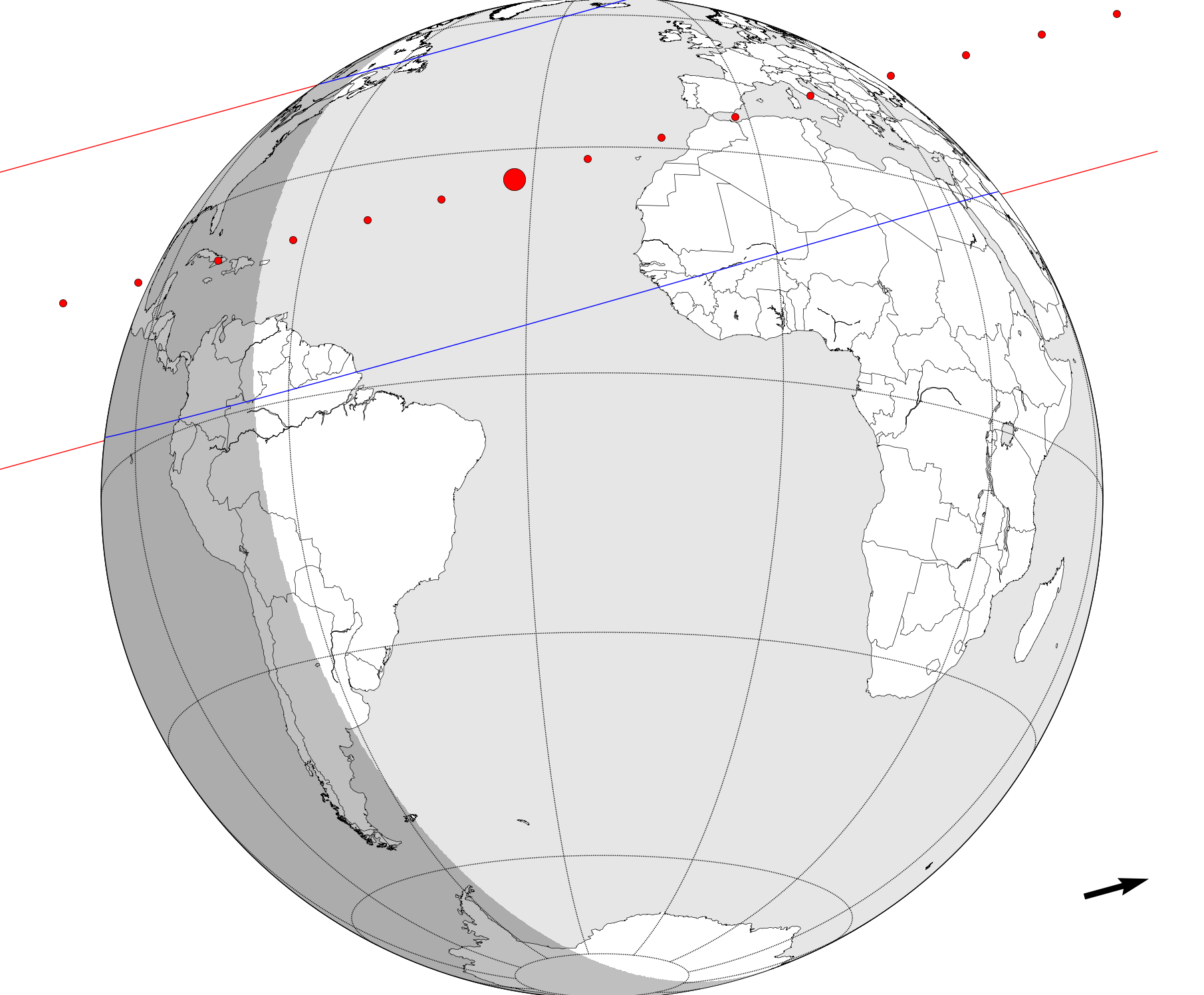}
\caption{Occultation of a mag. G 5.8 star by Io on 2021 April 2, 10:24 UTC. The predicted relative velocity of the event is 16.5 km/s.}
\label{Fig:I3}
\end{figure}               

\begin{figure}[h]
\includegraphics[width=0.45\textwidth]{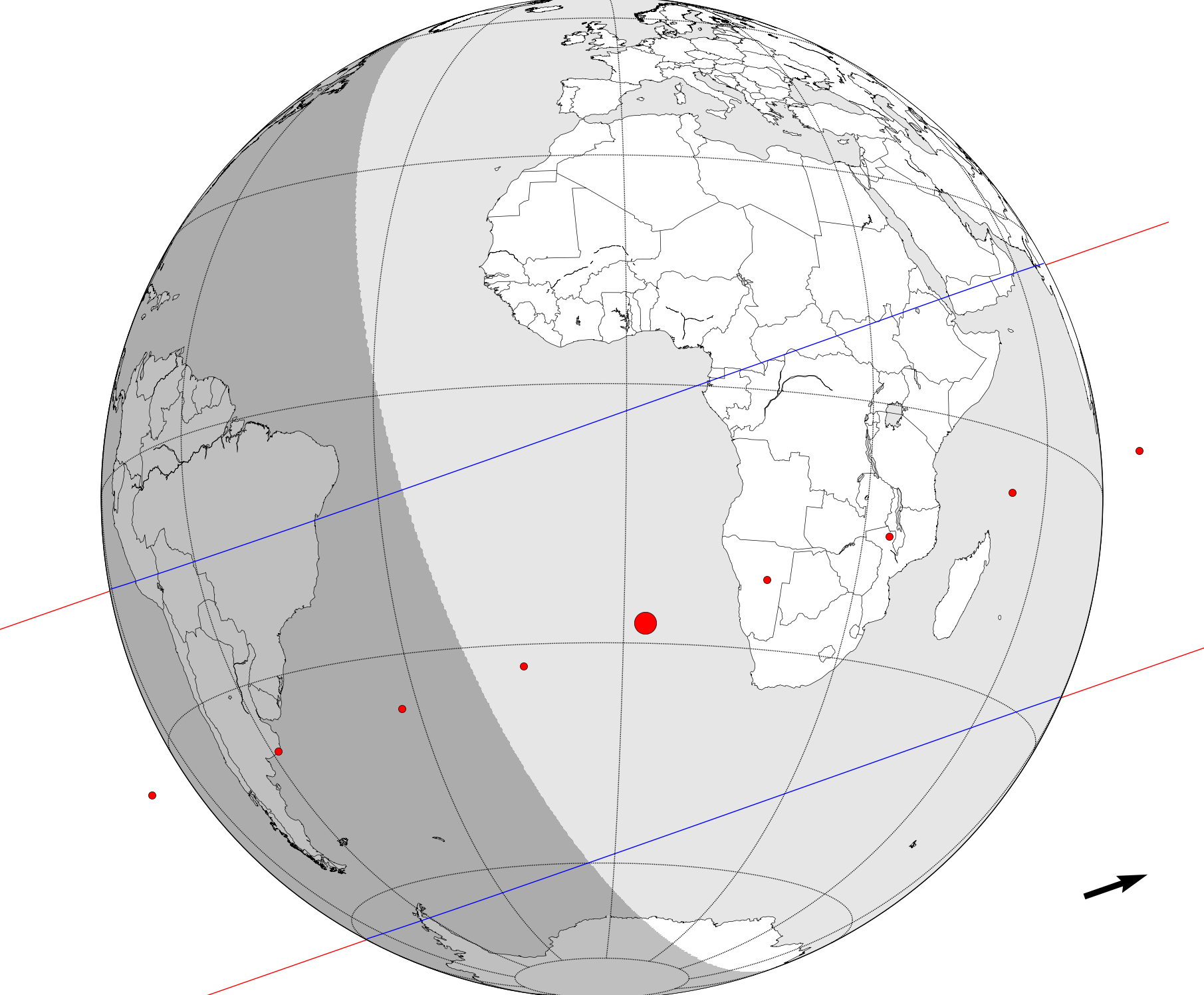}
\caption{Occultation of a mag. G 11.1 star by Ganymede on 2021 April 25, 07:55 UTC. The predicted relative velocity of the event is 27.7 km/s.}
\label{Fig:G1}
\end{figure}               

\begin{figure}[h]
\includegraphics[width=0.45\textwidth]{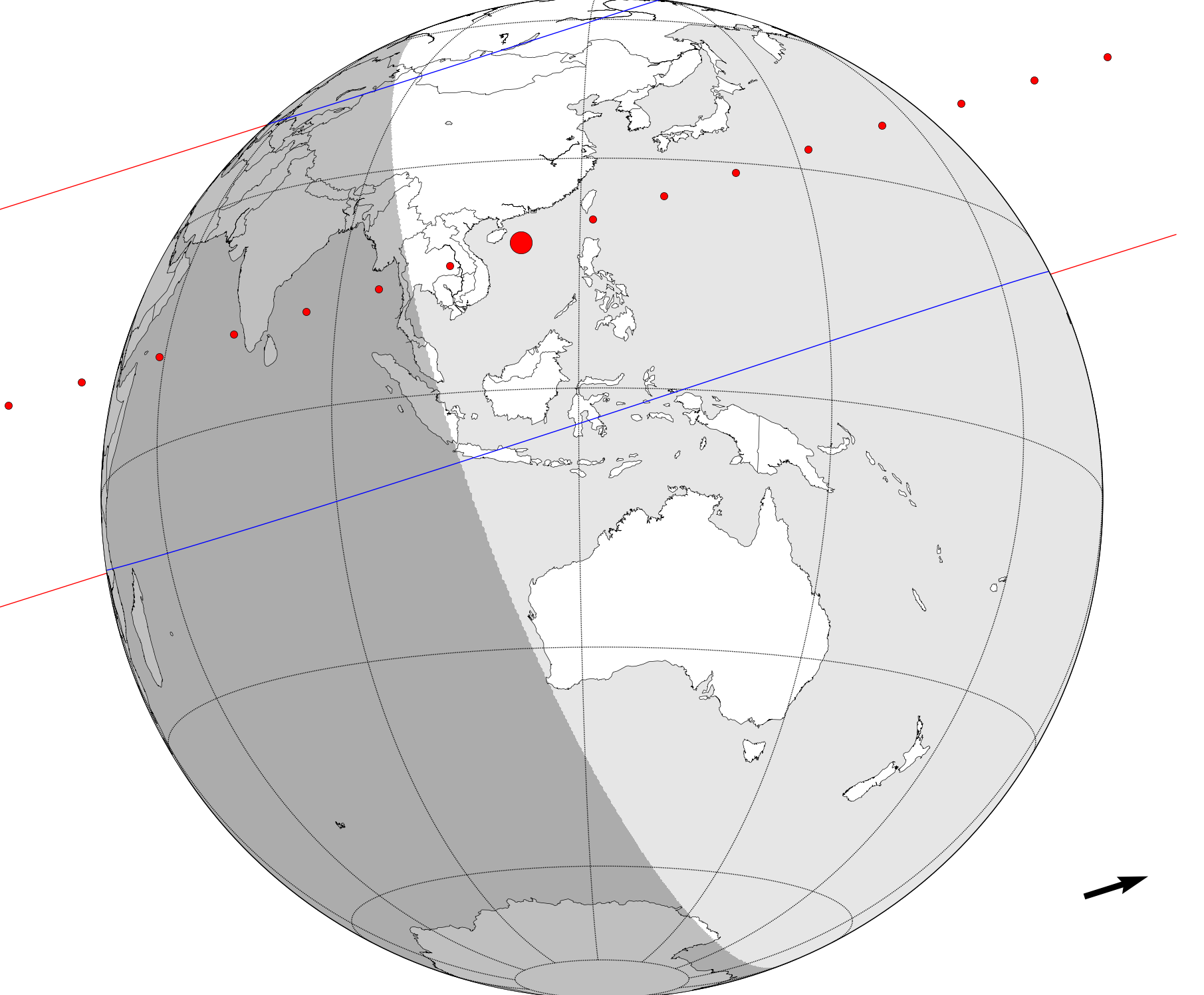}
\caption{Occultation of a mag. G 10.4 star by Callisto on 2021 May 4, 23:01 UTC. The predicted relative velocity of the event is 16.3 km/s.}
\label{Fig:C3}
\end{figure}               

\end{appendix}

\end{document}